\documentclass[a4paper,12pt]{aa}

\usepackage{graphicx}
\usepackage{amsmath}
\usepackage{amssymb}
\usepackage{xspace}
\usepackage{tikz}
\usepackage{txfonts}
\usetikzlibrary{arrows,positioning,shapes,optics}
\usepackage{siunitx}
\colorlet{papercolor}{red!60!black}
\tikzset{
    >=stealth,
    nist/.style={
      text=gray,
      rectangle,
      rounded corners,
      draw=gray, thick,
      text width=6.5em,
      minimum height=2em,
      text centered},
    paper/.style={
      rectangle,
      rounded corners,
      draw=papercolor, very thick,
      text width=6.5em,
      minimum height=2em,
      text centered,
      text = papercolor},
    stardice/.style={
           rectangle,
           rounded corners,
           draw=black, thick,
           text width=6.5em,
           minimum height=2em,
           text centered},
    powers/.style={
      font=\scriptsize,
      text centered},
    atnist/.style={
      <->,
      thick,
      shorten <=2pt,
      shorten >=2pt,
      color=gray,
    },
    inthepaper/.style={
      <->,
      color=papercolor,
      very thick,
      shorten <=2pt,
      shorten >=2pt,},
    atlpnhe/.style={
      <->,
      thick,
      shorten <=2pt,
      shorten >=2pt,}
}

\DeclareMathOperator{\E}{\mathbb{E}}
\DeclareMathOperator{\var}{var}
\DeclareMathOperator{\erf}{erf}
\newcommand\pixcount[1][p,i]{\ensuremath{q_{#1}}\xspace}

\title{StarDICE I: sensor calibration bench and absolute photometric
  calibration of a Sony IMX411 sensor}

\author{
Marc Betoule\inst{1}
 \and Sarah Antier\inst{4}
 \and Emmanuel Bertin\inst{5}
 \and Pierre Éric Blanc\inst{6}
 \and Sébastien Bongard\inst{1}
 \and Johann Cohen Tanugi\inst{7,10}
 \and Sylvie Dagoret-Campagne\inst{2}
 \and Fabrice Feinstein\inst{3}
 \and Delphine Hardin\inst{1}
 \and Claire Juramy\inst{1}
 \and Laurent Le Guillou\inst{1}
 \and Auguste Le Van Suu\inst{6}
 \and Marc Moniez\inst{2}
 \and Jérémy Neveu\inst{2,11}
 \and Éric Nuss\inst{7}
 \and Bertrand Plez\inst{7}
 \and Nicolas Regnault\inst{1}
 \and Eduardo Sepulveda\inst{1}
 \and Kélian Sommer\inst{7}
 \and Thierry Souverin\inst{1}
 \and Xiao Feng Wang\inst{8,9}
}
\institute{
LPNHE, CNRS/IN2P3 \& Sorbonne Université, 4 place Jussieu, 75005 Paris, France
 \and Universit\'e Paris-Saclay, CNRS, IJCLab, 91405, Orsay, France
 \and Aix Marseille Univ, CNRS/IN2P3, CPPM, Marseille, France
 \and Artemis, Observatoire de la Côte d’Azur, Université Côte d’Azur, Boulevard de l’Observatoire, 06304 Nice, France
 \and Sorbonne Université, IAP, Paris, F-75014, France
 \and Université d’Aix-Marseille \& CNRS, Observatoire de Haute-Provence, 04870 Saint Michel l’Observatoire, France
 \and LUPM, Université Montpellier \& CNRS, F-34095 Montpellier, France
 \and Physics Department and Tsinghua Center for Astrophysics, Tsinghua University, Beijing, 100084, China
 \and Beijing Planetarium, Beijing Academy of Science and Technology, Beijing, 100044, China
 \and LPC, université Clermont Auvergne, CNRS, F-63000 Clermont-Ferrand, France
 \and Sorbonne Universit\'e, CNRS, Universit\'e de Paris, LPNHE, 75252 Paris Cedex 05, France
}

\begin{document}
\abstract{The Hubble diagram of type-Ia supernovae (SNe-Ia) provides
  cosmological constraints on the nature of dark energy with an
  accuracy limited by the flux calibration of currently available
  spectrophotometric standards. This motivates new developments to
  improve the link between existing astrophysical flux standards and
  laboratory standards.}
{The StarDICE experiment aims at establishing a 5-stage metrology
  chain from NIST photodiodes to stars, with a targeted accuracy of
  \SI{1}{mmag} in $griz$ colors. We present the first two stages,
  resulting in the calibration transfer from NIST photodiodes to a
  demonstration \SI{150}{Mpixel} CMOS sensor (Sony IMX411ALR as implemented
  in the QHY411M camera by QHYCCD). As a side-product, we provide full
  characterization of this camera which we believe of potential
  interest in astronomical imaging and photometry and specifically
  discuss its use in the context of gravitational wave optical
  follow-up.}
{A fully automated spectrophotometric bench is built to perform the
  calibration transfer. The sensor readout electronics is studied
  using thousands of flat-field images from which we derive stability,
  high resolution photon transfer curves and estimates of the
  individual pixel gain. The sensor quantum efficiency is then
  measured relative to a NIST-calibrated photodiode, in a well defined
  monochromatic light-beam from $375$ to \SI{1078}{\nano\metre}. Last,
  flat-field scans at 16 different wavelengths are used to build maps
  of the sensor response, fully characterizing the sensor for absolute
  photometric measurements.}
{We demonstrate statistical uncertainty on quantum efficiency below
  \SI{0.001}{e^-/\gamma} between \SI{387}{nm} and \SI{950}{nm}, the
  range being limited by the sensitivity decline of the tested sensor
  in the infrared. Systematic uncertainties in the bench optics are
  controlled at the level of \SI{1e-3}{e^-/\gamma}. Linearity issues
  are detected at the level of \SI{5e-3}{e^-/\gamma} for the tested
  camera and require further developments to fully
  correct. Uncertainty in the overall normalization of the QE curve
  (without relevance for the cosmology, but relevant to evaluate the
  performance of the camera itself) is 1\%. Regarding the camera we
  demonstrate stability in steady state conditions at the level of
  \SI{32.5}{ppm}. Homogeneity in the response is below
  \SI{1}{\percent} RMS across the entire sensor area. Quantum
  efficiency stays above \SI{50}{\percent} in most of the visible
  range, peaking well above \SI{80}{\percent} between \SI{440}{nm} and
  \SI{570}{nm}. Differential non-linearities at the level of
  \SI{1}{\percent} are detected. A simple 2-parameter model is
  proposed to mitigate the effect and found to adequately correct the
  shape of the PTC on half the numerical scale. No significant
  deviations from integral linearity were detected in our limited
  test. Static and dynamical correlations between pixels are low,
  making the device likely suitable for galaxy shape measurements.}
{}

\maketitle

\section{Introduction}
\label{sec:introduction}

The calibration of wide-field optical surveys is a subject of active
research, driven by requirements from the use of type Ia supernovae to
measure the evolution of luminosity distance with redshift (the Hubble
diagram), one of the main probes of dark energy. Essentially, this
measurement involves the comparison of the apparent fluxes of
supernovae at different redshifts. Errors in the differential flux
calibration between the bluer photometric bands, in which the low
redshift events are observed, and the redder bands, in which high
redshift events fall, translate directly into a systematic error in
the Hubble diagram. In recent studies, the contribution of the
calibration uncertainty to the total uncertainty on the dark energy
equation of state parameter has been brought down to a level
comparable to the statistical uncertainty
\citep{2014A&A...568A..22B,2018ApJ...859..101S,2019ApJ...881...19J}. The
next generation of instruments, in particular the Vera C Rubin
observatory, requires a substantial step-forward in terms of
photometric accuracy to benefit from the one-order-of-magnitude
increase in the available statistics.

StarDICE is one of the experiments aiming at establishing a metrology
chain between laboratory flux references (silicon photodiodes
calibrated by NIST) and stars from the CALSPEC library of
spectrophotometric standards \citep{2020AJ....160...21B}. As supernova
surveys are calibrated relative to these standard stars, establishing
this chain with sufficient accuracy essentially solves the calibration
issue of the Hubble diagram.

With dark currents of the order of \SI{100}{\femto\ampere} however,
the reference photodiodes cannot be used reliably to measure
irradiances fainter than \SI{1e-11}{W\per\centi\meter\squared} and
intermediate steps are required to bridge the remaining gap
leading to the irradiance of mag 13 stars of the order of
\SI{1e-19}{W\per\nano\meter\per\centi\meter\squared}. Experiments vary
in the way they achieve and control the flux reduction between the
calibration reference and the target instrument. Examples of proposed
strategies are diffusion on a screen
\citep{2007ASPC..364..373S,2010ApJS..191..376S,2012ApJ...750...99T,2013arXiv1302.5720M,2020SPIE11447E..5UF},
and diffusion in a sphere of short light pulses
\citep{2018SPIE10704E..20C}. The most advanced experiment to date
\citep{2017A&A...607A.113L,Kusters2019Improving,2020SPIE11447E..71K}
uses a combination of diffusion in integrating spheres together with
extremely sensitive calibrated photodiodes \citep{lira}.

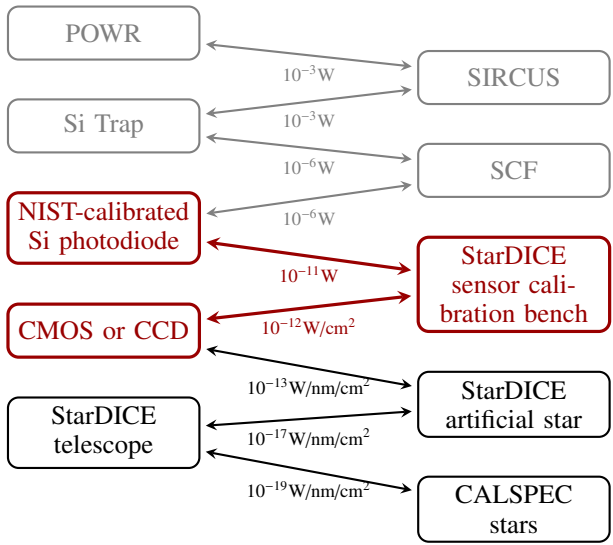
\begin{figure}
  \centering
\begin{tikzpicture}[node distance=1cm, auto,]
  \node[nist] (POWR) {POWR};
  \node[nist, below=0.5cm of POWR] (sitrap) {Si Trap};
  \node[right=4cm of POWR] (dummy) {};
  \node[nist, below=0.1cm of dummy] (SIRCUS) {SIRCUS}
  edge[atnist] node[powers,midway,below]{$10^{-3} \text{W}$} (POWR)
  edge[atnist] node[powers,midway,below]{$10^{-3} \text{W}$} (sitrap);
  \node[paper, below=0.5cm of sitrap] (siphot) {NIST-calibrated Si photodiode};
  \node[nist, below=0.5cm of SIRCUS] (SCF) {SCF}
  edge[atnist] node[powers,midway,below]{$10^{-6} \text{W}$} (siphot)
  edge[atnist] node[powers,midway,below]{$10^{-6} \text{W}$} (sitrap);
  \node[paper, below=0.5cm of siphot] (CCD) {CMOS or CCD};
  \node[paper, below=0.5cm of SCF] (ardice) {StarDICE sensor calibration bench}
  edge[inthepaper] node[powers,midway,below] {$10^{-11} \text{W}$} (siphot)
  edge[inthepaper] node[powers,midway,below] {$10^{-12} \text{W/cm}^2$} (CCD);
  \node[stardice, below=0.5cm of CCD] (OT0) {StarDICE telescope};
  \node[stardice, below=0.5cm of ardice] (dice) {StarDICE artificial star}
  edge[atlpnhe] node[powers,midway,below] {$10^{-17} \text{W/nm/cm}^2$} (OT0)
  edge[atlpnhe] node[powers,midway,below] {$10^{-13}\text{W/nm/cm}^2$} (CCD);
    \node[stardice, below=0.5cm of dice] (CALSPEC) {CALSPEC stars}
  edge[atlpnhe] node[powers,midway,below] {$10^{-19}\text{W/nm/cm}^2$} (OT0);
\end{tikzpicture}
\caption{Metrology chain of the StarDICE experiment, with
    light detectors in the left column and light sources in the right
    column. Each arrow represents a step in the chain and the label
    gives the order of magnitude of the beam intensity, irradiance or
    spectral irradiance depending if the beam is contained or extended
    and monochromatic or not. The steps in gray are conducted at NIST
    \citep{houston2008detectors} and result in a silicon photodiode
    calibrated against an electrical substitution cryogenic
    radiometer. The two steps in red are covered in this paper and
    provide calibrated sensors with a much lower dark current than
    the photodiode. The rest of the chain leading to astrophysical
    standards will be the subject of forthcoming
    papers.}\label{fig:metrology}
\end{figure}

The StarDICE proposal consists of a five-stage chain depicted in
Fig.~\ref{fig:metrology}, which relies on the near-field calibration
of an ultra-faint (less than \SI{1}{\micro\W} of optical power) and
stable light source which in turn serves as a distant (\SI{\sim
  100}{m}) in-situ calibration reference for a small astronomical
telescope. To achieve this task, sensitive calibrated devices, either
cooled CMOS or CCDs, are required to rapidly and accurately map the
radiant intensity of this artificial star at short distance
(\SI{\sim20}{cm}). The present paper, the first in a series describing
the chain, details the setup of a spectrophotometric test bench built
to transfer the NIST photodiode calibration to the nearby calibration
sensor with the required \SI{\sim0.1}{\percent} accuracy, and
demonstrating the first two steps from the chain.

The choice of a suitable sensor deserves some discussion. Since the
early 2000s, many applications of image sensors have gradually shifted
from charged coupled devices (CCD) to complementary metal oxide
semiconductor image sensors (CMOS image sensors or CIS). In the field
of astronomy, however, the important characteristics are quantum
efficiency (especially in the near infrared), uniformity, and
linearity in the response (provided simple corrections such as
flat-fielding), and sensitive area. In the case of these
characteristics, the field has been well served by the developments in
CCD technology such as infrared sensitive ultra-thick substrates,
multi-layer coatings, 4-side buttable sensors, and increase in the
number of output amplifiers. A recent example is given by the Teledyne
e2v CCD250-82 sensor developed for the camera of the Vera Rubin
Observatory \citep{2014SPIE.9154E..1PJ,2016SPIE.9915E..0XO}, which
combines all these recent developments. As a consequence, professional
astronomy experiments still mainly rely on CCDs. Nevertheless, the
last decade has seen increased availability of QE-boosting
technologies such as backside illumination (BSI) in CIS, improvements
in the performance and arrangement of column-parallel ADCs allowing
larger area and 3-side buttable sensors. Specific applications are
already making use of the advantages of CIS such as the ability to
address subsets of individual pixels as in the TAOS-II project
\citep{2021PASP..133c4503H}, which uses the Teledyne CIS113
\citep{10.1117/12.2561204} with deported ADCs. Another useful
characteristic of CIS is the achievement of high frame rates coupled
with low readout noise ($\lesssim$\SI{1}{e^-}), thanks to highly
parallel readout. This feature has important potential applications in
high-resolution ground-based astronomy which requires short exposure
times (\SI{\sim10}{\milli\second}) to freeze the speckle pattern
caused by atmospheric turbulence.

In this first study we characterize a camera equipped with the
recently available \SI{150}{Mpixel} IMX411ALR Sony CMOS sensor as implemented
in the QHY411M camera. With a large \SI{53 x 40}{\milli\metre}
sensitive area, low dark current without the need of extensive cooling
and negligible read-out noise, this sensor appears very convenient for
the direct mapping of low irradiance we are pursuing.

The interest in this camera also extends well beyond the StarDICE
experiment. The sampling of the large area by small
\SI{3.76}{\micro\metre} side pixels makes the sensor well suited for
wide-field 1m-class telescopes. This kind of instrument is
particularly useful in the follow-up of gravitational waves and
neutrino events or as survey instrument for the serendipitous
discovery of optical transients such as supernovae or kilonovae. This
field of research requires a monitoring of multiple sources, rapidly
fading and poorly localized (\SI{> 1}{\deg\squared}
\citealt{2018LRR....21....3A}). The Global Rapid Advanced Network
Devoted to Multi-messenger Addicts (GRANDMA,
\citealt{2020MNRAS.497.5518A,2022MNRAS.515.6007A}) needs telescopes
with photometric performance similar to the Zwicky Transient Facility
\citep{Bellm_2018} (which has a magnitude limit of 21 for \SI{30}{s}
exposures in $r$ band) at a significantly lower cost to be replicated
at different sites. It is particularly interesting to obtain high
efficiency in the near infrared band since it allows the imaging of
the peak of the red kilonova component even if the blue is too weak to
be detected by 1m-class telescopes and alert the sensitive
spectrographs
\citep{Metzger:2019zeh,2019PASP..131f8004A,LSC_MM_2017ApJ}. The
unknown feature here is the absolute quantum efficiency of the camera
over the required wavelength range. A secondary aim of the paper is
therefore to study the camera and measure its quantum efficiency as a
reference for the GRANDMA collaboration or other potential
applications in astronomy.

The rest of the paper is organised as follows.
Section~\ref{sec:instrumental-setup} provides a detailed description
of the instrumental setup. Section~\ref{sec:dataset} gives an overview
of the measurement campaign and describes in detail the measurement
process. The main results and associated uncertainties are discussed
in Section~\ref{sec:results}. The results are summarized and discussed
in section~\ref{sec:discussion}.

\section{Description of the sensor calibration bench}
\label{sec:instrumental-setup}

The StarDICE sensor calibration bench is designed to calibrate CCD or
CMOS cameras relative to a flux standard established by
NIST\citep{houston2008detectors}. The light illumination system has
been designed to provide both flat-fielding and mono-chromatic local
illumination capabilities in order to fully map the camera quantum
efficiency and readout gain.

The camera tested in this study is a QHY411M equipped with a
back-illuminated monochrome CMOS sensor IMX411ALR. The sensor
illuminated area consists of \SI{10654 x 14206}{pixels}
(\SI{151.3}{\mega pixels}) with a pixel width of
\SI{3.76}{\micro\meter} and a saturation capacity around
\SI{80}{ke^-}. The sensor has a rolling electronic
shutter\footnote{See e.g. \citet{Pace2021} for a description of CMOS
  image sensors shuttering.} and its ADC array allows 16 bit readout
of the pixel array at a maximum frame rate of \SI{2} {image/s}, with
adjustable gain. The camera thermoelectrically cools the sensor down
to \SI{-45}{\celsius} (adjustable) and the generated heat is extracted
by a closed-loop circulation of water at \SI{16}{\celsius}. The QHY411
camera is connected to the control computer through a USB3 interface,
and commands are sent through a python wrapper on top of the QHY
SDK.\footnote{\url{https://github.com/ebertin/qhpyccd}} The data were
acquired in exposure mode, not in video mode, which we did not manage
to get working with from Linux. The tunable electronic bias and
amplifier gain of the camera were kept fixed to their respective
values of \num{100} and \num{60}, which delivered a gain on the order
of \SI{1.2}{e^-/ADU}, effectively mapping the sensor
dynamics\footnote{The single pixel full well capacity is expected to
  be \SI{80}{\kilo e^-} according to the manufacturer
  website. \url{https://www.qhyccd.com/qhy411-qhy461/}.}  over the 16
bits.

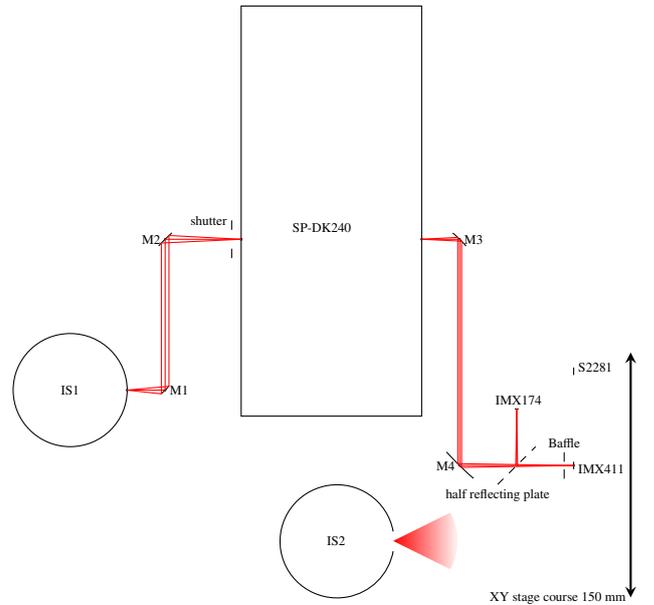
\begin{figure}
  \centering
  \begin{tikzpicture}[use optics, scale=0.25, every node/.style={scale=0.5},
    monobeam/.style={red}]
    \usetikzlibrary{intersections}
    \draw (0, 0) circle (1pt);
    \pgfmathsetmacro{\MONEY}{-8};
    \pgfmathsetmacro{\MONEYcm}{-8cm};
    \pgfmathsetmacro{\MTWOY}{2cm};
    \pgfmathsetmacro{\MFOURY}{-12};
    \pgfmathsetmacro{\MFOURYcm}{-12cm};
    \begin{scope}[yshift=\MONEYcm, xshift=-9cm]
    \draw (0,0) circle (3cm);
    \node (0, 0) {IS1};
    \end{scope}
    
    \begin{scope}[yshift=\MONEYcm-1cm, xshift=-6cm, xscale=-1]
    \pgfmathsetmacro{\fone}{1};
    \pgfmathsetmacro{\Done}{0.5};
    \begin{scope}
    \clip (-2*\fone, \fone) circle (0.5*\Done cm);
    \draw (-2*\fone, \fone) circle (1pt);
    \draw (-3*\fone, 9/4*\fone) parabola bend (0, 0) (3*\fone,9/4*\fone);
    \end{scope}
    \draw (-2*\fone, \fone) node[right] {M1};
    \draw (0, \fone) circle (1pt);
    \end{scope}

    \begin{scope}[yshift=\MTWOY, xshift=0cm]
    \pgfmathsetmacro{\fone}{-2};
    \pgfmathsetmacro{\Done}{+1};
    \begin{scope}
    \clip (2*\fone, \fone) circle (0.5*\Done cm);
    \draw (2*\fone, \fone) circle (1pt);
    \draw (-3*\fone, 9/4*\fone) parabola bend (0, 0) (3*\fone,9/4*\fone);
    \end{scope}
    \draw (2*\fone, \fone) node[left] {M2};
    \draw (0, \fone) circle (1pt);
    \end{scope}

    \draw (-.5, -1) -- (-0.5, -0.5);
    \draw (-.5, 0.5) -- (-0.5, 1) node[left] {shutter};
    \begin{scope}[yshift=-9.38cm]
    \draw (0, 0) rectangle (9.5, 21.75);
    \draw (4.25, 10) node {SP-DK240};
    \end{scope}

    \begin{scope}[yshift=\MTWOY-1cm, xshift=9.5cm, xscale=-1]
    \pgfmathsetmacro{\fone}{-1};
    \pgfmathsetmacro{\Done}{+1};
    \begin{scope}
    \clip (2*\fone, \fone) circle (0.5*\Done cm);
    \draw (2*\fone, \fone) circle (1pt);
    \draw (-3*\fone, 9/4*\fone) parabola bend (0, 0) (3*\fone,9/4*\fone);
    \end{scope}
    \draw (2*\fone, \fone) node[right] {M3};
    \draw (0, \fone) circle (1pt);
    \end{scope}

    \begin{scope}[yshift=\MFOURYcm-3cm, xshift=9.5cm+8cm, xscale=-1]
    \pgfmathsetmacro{\fone}{3};
    \pgfmathsetmacro{\Done}{+2};
    \begin{scope}
    \clip (2*\fone, \fone) circle (0.5*\Done cm);
    \draw (2*\fone, \fone) circle (1pt);
    \draw (-3*\fone, 9/4*\fone) parabola bend (0, 0) (3*\fone,9/4*\fone);
    \end{scope}
    \draw (0, \fone) circle (1pt);
    \draw (2*\fone, \fone) node[left] {M4};
    \end{scope}

    \draw (9.5 + 8, \MFOURY - 0.2+5) -- (9.5 + 8, \MFOURY + 0.2+5) node[right] (S2281) {S2281};
    \draw (9.5 + 8, \MFOURY + 0.2) -- (9.5 + 8, \MFOURY - 0.2) node[right] (IMX411) {IMX411};
    \draw [<->, thick] (9.5 + 11, \MFOURY + 6) -- (9.5 + 11, \MFOURY-7) node[left] {XY stage course 150 mm};

    \draw (9.5 + 7.5, \MFOURY + 0.2) -- (9.5 + 7.5, \MFOURY + 0.7) node[above] (STOP) {Baffle};
    \draw (9.5 + 7.5, \MFOURY - 0.2) -- (9.5 + 7.5, \MFOURY - 0.7);
    
    \draw[dashed] (12.5+2 - 1 , -12 - 1) node[below] {half reflecting plate} -- (12.5+2 + 1 , -12+1);
    \draw (12.5+2-0.1, \MFOURY+3) -- (12.5+2+0.1, \MFOURY+3) node[above] {IMX174};
    \draw[monobeam] (-6, \MONEY) -- (-4, \MONEY) -- (-4, 0) -- (0, 0);
    \draw[monobeam] (-6, \MONEY) -- (-4-0.2, \MONEY-0.2) -- (-4-0.2, -0.2) -- (0, 0);
    \draw[monobeam] (-6, \MONEY) -- (-4+0.2, \MONEY+0.2) -- (-4+0.2, +0.2) -- (0, 0);
    \draw[monobeam] (9.5, 0) -- (9.5+2, 0) -- (9.5+2, \MFOURY) -- (9.5 + 8, \MFOURY);
    \draw[monobeam] (9.5, 0) -- (9.5+2-0.1, 0.1) -- (9.5+2-0.1, \MFOURY+0.1) -- (9.5 + 8, \MFOURY);
    \draw[monobeam] (9.5, 0) -- (9.5+2+0.1, -0.1) -- (9.5+2+0.1, \MFOURY-0.1) -- (9.5 + 8, \MFOURY);
    \draw[monobeam] (12.5+2, \MFOURY) -- (12.5+2, \MFOURY+3);
    \draw[monobeam] (12.5+2-0.05, \MFOURY+0.05) -- (12.5 + 2, \MFOURY+3);
    \draw[monobeam] (12.5+2+0.05, \MFOURY-0.05) -- (12.5+2, \MFOURY+3);

    \begin{scope}[yshift=-16cm, xshift=5cm]
    \draw (3,0.5) arc (10:350:3cm);
    \node (0, 0) {IS2};
    \shade[left color=red, right color=white] (3,0) -- (6, -1.5) arc (-30:30:3cm) -- (3,0);
    
    \end{scope}

  \end{tikzpicture}
  
  \caption{Sketch of the optics of the StarDICE sensor calibration
    bench. The characteristics of the optical elements are given in
    Table~\ref{tab:focaldiameter}. The light provided by the
    integrating sphere IS1 is filtered by the SP-DK240 monochromator
    and focused in a plane where the reference (S2281) and target
    (IMX411) sensors can be positioned thanks to a motorized
    XY-stage. An aperture stop (Baffle) is used to clean the focused
    beam from stray light. The half-reflecting plate redirects half of
    the light beam on a time-stability monitoring camera (IMX174). The
    linear stage can alternatively bring the sensors in direct view of
    the second sphere (IS2) providing flat-field illumination. }
  \label{fig:setup}
\end{figure}

The test camera is mounted on a two-axis linear stage in an optical
test bench where it can intercept either a wavelength tunable f/10
monochromatic light beam or a flat-field illumination beam. Other
light sensors, in particular a NIST calibrated Hamamatsu S2281
photodiode, are mounted in the same plane and can in turn intercept
the same beams. A sketch description of the setup is given in
Figure \ref{fig:setup}.

Intercalibration between detectors is obtained by taking the ratio of their
response to illumination by the same monochromatic
light-beam. Measurements of the detector uniformity and photon
transfer curves (PTC) are obtained from flat-field exposures in the second
beam.

\subsection{Monochromatic beam}
\label{sec:monochromaticbeam}

The monochromatic light beam is obtained by imaging the horizontal
output slit of an integrating sphere (IS1) on the vertical entrance
slit of a Czerny-Turner monochromator (SP-DK240) using a relay of two
90° off-axis parabolic mirrors (M1 and M2). The output slit of the
monochromator is then imaged on an aperture stop in front of the
detector plane by a second relay (M3 and M4). The focal length of the
monochromator is \SI{240}{\milli\metre} and its turret hosts three
\SI{68 x 68}{\milli\metre} gratings, accepting input beams with a
recommended minimal f-number of 3.9. The diameter and focal length of
the off-axis parabolic mirrors (OAP) are given in
Table~\ref{tab:focaldiameter}. The first relay shapes an $f/4$ input
beam which is then extended to $f/12$ by the second relay. The use of
mirrors makes the position of the focus point and image shape
conveniently achromatic.
\begin{table}
  \centering
  \caption{Diameter and focal length of the OAP mirrors.}
  \label{tab:focaldiameter}
  \begin{tabular}{ccc}
    \hline
    \hline
    Element & diameter & focal length \\
    & [mm (in)] & [mm (in)] \\
    \hline
    M1 & $12.7 (1/2)$ & $25.4 (1)$ \\
    M2 & $25.4 (1) $ & $50.8 (2)$ \\
    M3 & $25.4 (1) $ & $25.4 (1)$ \\
    M4 & $50.8 (2) $ & $76.2 (3)$ \\
    \hline
  \end{tabular}
\end{table}

Illumination in IS1 is provided by an electronic board in the
\SI{3}{inch} port featuring 48 LEDs at different central
wavelengths. The flux level of each LED is tunable by
software. Typically only one is shining at any given time to minimize
pollution of the output beam by out-of-band light. A switchable Ar-Hg
arc lamp (not shown in Fig. \ref{fig:setup}) is fitted to the
remaining port of the sphere, and is used to determine the wavelength
calibration and resolution of the monochromator.

\begin{figure*}
  \centering
  \includegraphics{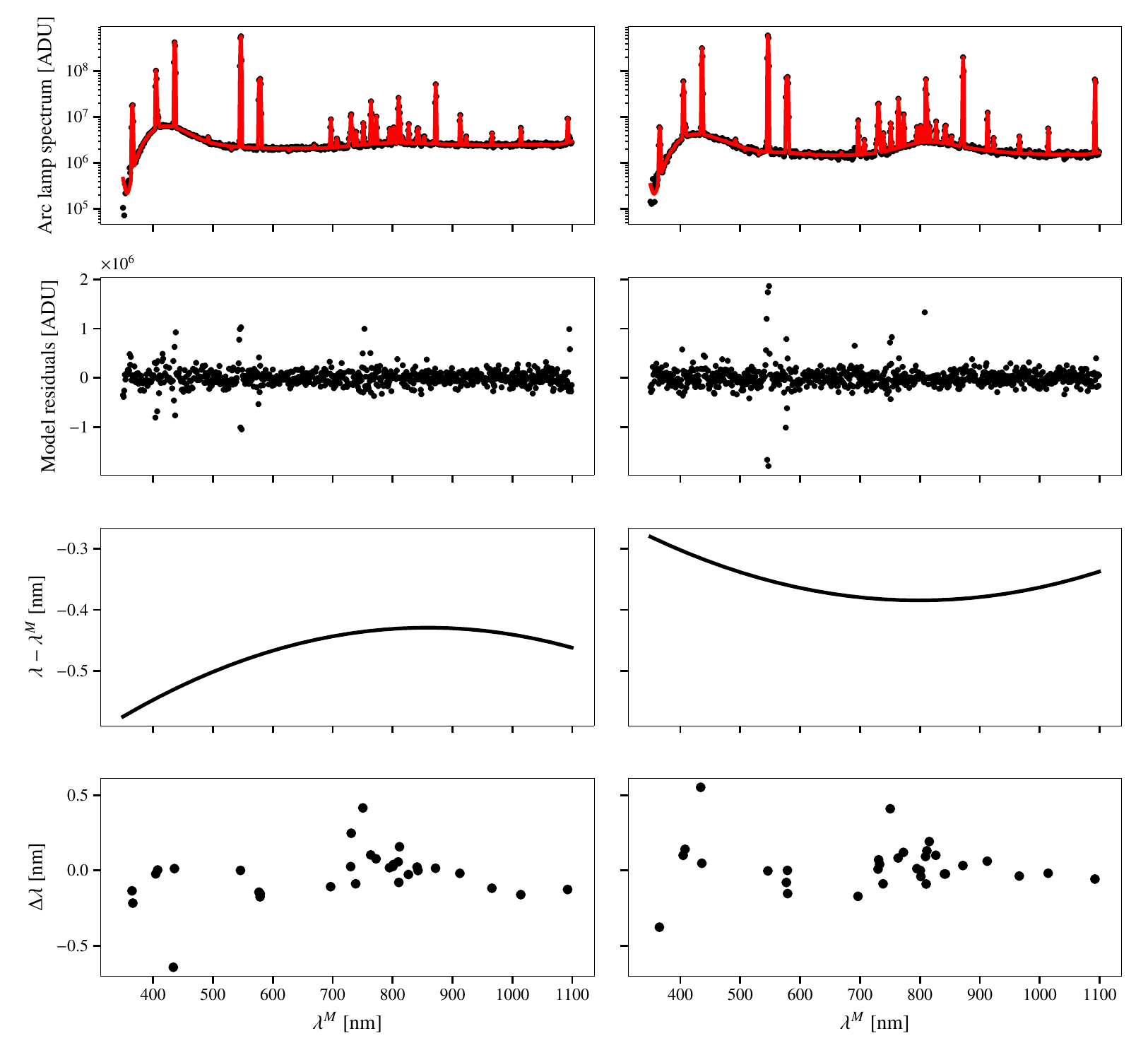}  
  \caption{Wavelength calibration model for the monochromatic
    calibration beam. Results for the blue-optimized grating in use
    for wavelength shorter than \SI{650}{\nano\metre} are presented in
    the left column, results for the infrared-optimized grating are
    presented in the right column. \emph{Top row:} Observed spectrum
    of the Hg-Ar arc lamp (black dots) and best-fit model (red
    curve). \emph{Second row:} Difference in the observed spectrum and
    the best-fit model. \emph{Third row:} Wavelength calibration
    component of the best fit model, that is the quantity to be added
    to the monochromator set wavelength ($\lambda^M$) to match the
    theoretical wavelength of the observed lines
    ($\lambda$). \emph{Last row:} Residuals to the wavelength
    calibration, that is difference in measured and theoretical line
    wavelength in the air remaining after wavelength calibration for
    the brightest lines in the lamp spectrum. The RMS in the
      difference is respectively \SI{0.17}{nm} and \SI{0.16}{nm} for
    the blue and red gratings.}
  \label{fig:wlcal}
\end{figure*}

Our target wavelength resolution is of the order of
\SI{2}{\nano\meter}, with gratings of \SI{1200}{g/\milli\metre} and a
Ebert angle of \SI{18.7}{\degree}. The theoretical dispersion relation
therefore varies between \SI{2.92}{\nano\metre/\milli\metre} and
\SI{3.47}{\nano\metre/\milli\metre} over the wavelength range
$1100$--\SI{300}{\nano\metre}. A slit width of \SI{635}{\micro\meter}
delivers a full width at half maximum (FWHM) ranging between
\SI{1.86}{nm} and \SI{2.2}{nm}. We keep both slits at this fixed width
across the entire range to avoid changing the image position and the
wavelength calibration. The vertical extent of the image is set by the
opening of IS1 output slits which is set around
\SI{500}{\micro\meter}. In this setting, the spatial extent of the
image is about \SI{2 x 3}{\mm} in the sensor plane. Due to the
relatively low wavelength resolution it is necessary to build a full
forward model of the measured spectrum of the Hg-Ar calibration arc lamp to
derive accurate wavelength calibration from unresolved doublets or
triplets of the lamp. The observed spectrum is modeled as a series of
lines of adjustable amplitude convolved with a triangular wavelength
response whose width is allowed to vary linearly with wavelength. The
wavelength calibration is adjusted as a second order polynomial of the
wavelength. The background light is developed on a 13-knots
B-spline. The model provides a fairly good description of the spectral
lamp data as illustrated in Fig.~\ref{fig:wlcal} for the two gratings
in use. The wavelength correction curves shown in the third row of the
figure are applied to all subsequent measurements and bring the
uncertainty on the wavelength calibration well below
\SI{1}{\nano\metre}.

The 50\%-reflective beam splitter inserted after the last mirror
redirects a fraction of the flux on a CMOS IMX174LLJ sensor which
monitors the time stability of the illumination for a given
wavelength.

To further clean the main beam from stray light and make the image
extension on the detector perfectly stable, the light is focused on a
fixed aperture stop slightly smaller than the image, behind which the
slowly diverging beam reaches the detectors.

Finally, a shutter at the entrance of the monochromator is used to
shut the beam off during measurements of the ambient and stray light
level.

\subsection{Flat-field illumination beam}
\label{sec:flatfieldbeam}

\begin{figure}
  \centering
  \includegraphics{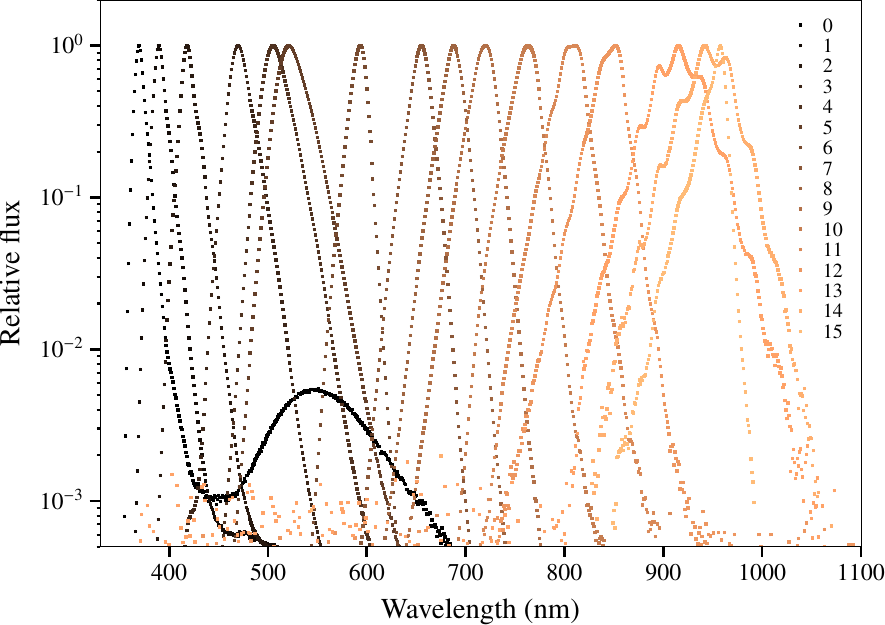}
  \caption{Normalized spectra of the 16 channels of the flat-field
    light source on a log scale. The UV LED displays some amount of
    fluorescence light, otherwise the wavelength range of each channel
    is quite narrow with respect to the typical width of broadband
    filters.}
  \label{fig:spectraflat}
\end{figure}

The flat-field light beam is provided by a second integrating sphere
(IS2) whose 1'' output port shines directly on the detectors. The
light is provided by 16 LEDs in a \SI{2}{inches} port without
wavelength filtering, so that the spectrum of the light emitted out of
the sphere corresponds roughly to the narrow spectrum of the shining
LED. A measurement of the output spectrum for each of the 16
independent channels is given in Figure~\ref{fig:spectraflat}. The
central wavelength ($\bar \lambda$) and spectral width (FWHM),
together with the wavelength of the flux maxima ($\lambda_{peak}$) in
all channels are provided in Table~\ref{tab:specraflat}. 

The intensity of the current flowing in each LED can be tuned
independently by software. The integrated surface brightness in the
sphere is monitored by a photodiode inserted in the last \SI{1}{inch}
port of the sphere. Over the full extent of the large IMX411 sensor,
the illumination varies by \SI{2.4}{\%} peak to valley, according to
the mapping of the beam reconstructed from the images taken during the
uniformity scan (see Sect.~\ref{sec:uniformityresults}).

\begin{table}
  \centering
  \caption{Central wavelength and spectral width in the flat-field light source.}
  \begin{tabular}{cccc}
    \hline
    \hline
    Channel & $\lambda_{peak}$ & \rule{0pt}{2.2ex}$\bar\lambda$ & FWHM \\
            & [nm] & [nm] & [nm] \\
    \hline
    0 & 370 & 381.4 & 8 \\
    1 & 390 & 392.4 & 10 \\
    2 & 418 & 421.3 & 11 \\
    3 & 470 & 473.2 & 19 \\
    4 & 505 & 509.8 & 25 \\
    5 & 521 & 526.7 & 30 \\
    6 & 593 & 591.2 & 15 \\
    7 & 656 & 653.6 & 19 \\
    8 & 688 & 686.2 & 19 \\
    9 & 720 & 716.9 & 23 \\
    10 & 763 & 759.9 & 24 \\
    11 & 811 & 804.6 & 27 \\
    12 & 852 & 843.4 & 28 \\
    13 & 915 & 915.3 & 52 \\
    14 & 942 & 952.2 & 42 \\
    15 & 958 & 956.3 & 16 \\
    \hline
  \end{tabular}
  \label{tab:specraflat}
\end{table}

\section{Dataset}
\label{sec:dataset}

Throughout this study we performed 5 different kinds of measurements:
\begin{enumerate}
\item \label{item:3} Dark current and readout noise measurements: the
  sensor is protected from any illumination by an aluminum front
  cover.
\item \label{sec:dataset-2} Stability measurements: the camera is kept
  fixed in front of the flat-field beam, at fixed illumination level
  and fixed exposure time. This experiment is conducted at 4 different
  (stabilized) temperatures to study temperature dependence.
\item \label{sec:dataset-1} Photon transfer curve and linearity
  measurements: Same configuration as dataset~\ref{sec:dataset-2}
  except that the illumination level is varied by changing either the
  source brightness or the exposure time. The sensor temperature is
  kept fixed at \SI{0}{\celsius} for this measurement and all the
  remaining ones.
\item \label{item:1} Uniformity measurements: hardware configuration
  similar to dataset~\ref{sec:dataset-2} except that the position of
  the sensor is varied, scanning the flat-field beam.
\item \label{item:2} Quantum efficiency measurements: interleaved
  measurements of the brightness of the monochromatic beam with the
  camera and the reference photodiode. The measurement is repeated
  scanning every nanometer in wavelength.
\end{enumerate}
We details the specifics of each measurement in the following, except
for the dark current measurement which is straightforward.

\subsection{Stability measurements}
\label{sec:stability}
\begin{figure}
  \centering
  \includegraphics{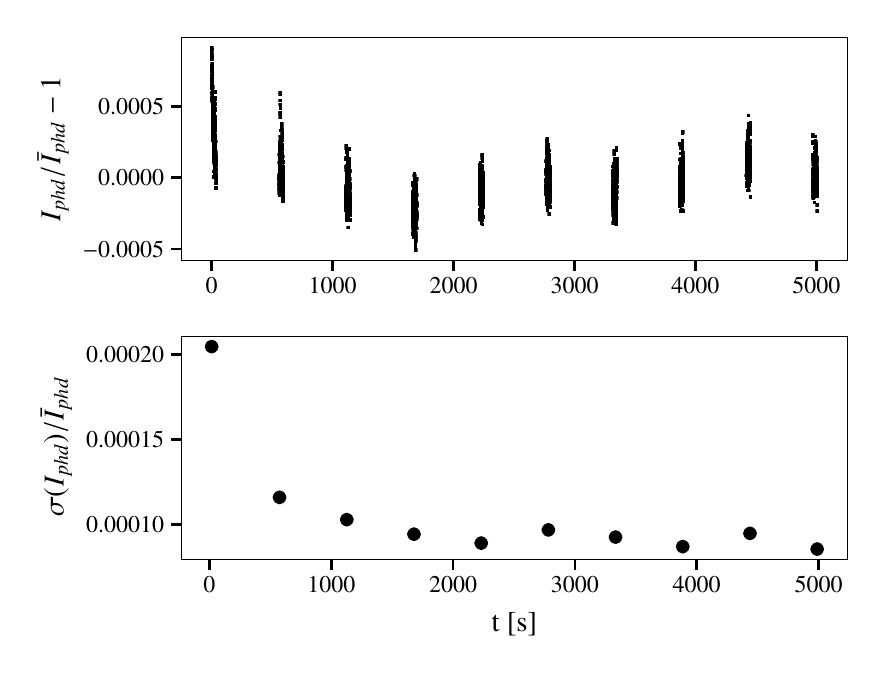}
  \caption{\emph{Top:} Evolution of the photocurrent delivered by the
    monitoring photodiode in the flat-field light source (IS2) with
    respect to its average over the entire period. \emph{Bottom:}
    Relative RMS of the photocurrent computed over 500 successive
    samples.}
  \label{fig:illuminationstability}
\end{figure}

A Large series of images were obtained with the camera roughly aligned
on the center of the flat-field beam to test the readout electronics.
The fluctuation of the illumination during a sequence at a given flux
level can be estimated from measurements with the monitoring
photodiode installed in IS2. The fluctuations in the illumination
typically drop below $10^{-4}$ RMS, after a warm-up period of
\SI{1000}{\s} as is shown in Fig.~\ref{fig:illuminationstability}.

The stability measurement is performed around \SI{510}{\nano\metre}
using LED channel 4. A stabilized current of \SI{10}{\milli\A} is
established in the LED and the thermoelectric cooling system of the
sensor is set to a target temperature. This was repeated at 4 set
points between $-5$ and \SI{10}{\celsius}. The exposure time is set to
\SI{500}{\milli\second} which results in an average pixel charge of
\SI{36}{\kilo e^-}. A first batch of $I=100$ exposures are taken to
wait for the cooling and stabilization of the system temperature. The
stability measurement is performed on a subsequent batch of $I=100$
exposures. Denoting \pixcount the count measured in pixel $p$ in image
number $i$, all images containing the same number of pixels $P$, the
following statistics can be gathered on the fly for each batch of
images:
\begin{eqnarray}
  \label{eq:stat}
  M_p &=& \frac1I\sum_i \pixcount\\
  V_p^\delta &=& \frac1I\sum_i \pixcount \pixcount[p+\delta, i]\\
  \label{eq:statlast}
  m_i &=& \frac1P\sum_p \pixcount
\end{eqnarray}
As described in appendix~\ref{sec:stat-model-phot}, the readout gain
can be reconstructed from the relation between the temporal mean of
individual pixels $M_p$ and the mean of the squares $V_p^0$, while the
mean of individual images $m_i$ tracks fluctuations in the
illumination level.

The individual images are not stored in this approach. Although the
absence of storage prevents the use of robust statistics, it is
preferred in this study due to the very large size of individual
images. Inter pixel products ($V_p^\delta$ for $\delta >0$) can be
stored to study inter-pixel covariances. This is however not needed
for the stability study. We keep track of the cooling power of the
sensor at the end of image $i$ and $I_{phd}(t)$ the photocurrent
delivered by the photodiode during the sequence.

\subsection{Photon Transfer Curve and linearity measurements}
\label{sec:ptc}

The measurement of the PTC is similar to that of the stability except
that the sensor temperature is now kept fixed at \SI{0}{\celsius}, and
the exposure time and LED current are varied. Below, we combine data
from 6 different sets whose characteristics are given in
Table~\ref{tab:ptc}.  Sets 1--3 are dedicated to a fine measurement of
the gain and linearity on the first third of the scale \SI{<15}{\kilo
  e^-} which covers the entire dynamic range of the QE measurement. To
mitigate potential deviations from Poissonian statistics due to the
brighter-fatter effect (see below), the statistics $M_p$ and $V_p$ are
computed on binned pixels, and divided by the number of pixels in the
bins to lie on the same ADU scale as the non-binned statistics. Sets
4--6 sample the full scale without binning.

Outside of the sensor calibration bench, we also gathered two
supplemental datasets using a simpler illumination source to enable
experimental study of correlations between neighboring pixels as a
signature for inter-pixel capacitance or brighter-fatter effects
(dataset A), and mapping of the individual pixel gain using a large
number of images (dataset B and C). In this configuration, the camera
was capped with a 3D-printed hollow cone holding a simple
\SI{450}{\nano\metre} LED behind a diffusive screen. Switching to an
external illumination source allowed to free the bench for the
calibration of another camera during these time-consuming
measurements. The illumination stability with this alternate source is
however one order of magnitude worse than what is achieved on the
bench. Therefore we do not rely on these supplemental datasets for our
baseline determination of the gain. For study A, in addition to the
statistics presented in the previous section, we also gather
$V_p^{01}$ and $V_p^{10}$, the mean of the cross-products between
immediately neighboring pixels along a line and along a column.
\begin{table}
  \centering
  \caption{Parameter range for the photon transfer curve and linearity
    datasets}
  \begin{tabular*}{\linewidth}{clcccc}
    \hline
    \hline
    Set & $I$ & $\tau_{exp}$ & $I_{led}$ & Binning & Statistics\\
    & & [s] & [mA] \\
    \hline
    1 & 10  & 0.3         &  10.0 -- 49.8 & $6\times6$ &$V^{0}$\\
    2 & 100 & 0.01 -- 1.0 &  0.0 -- 20.0  & $6\times6$ &$V^{0}$\\
    3 & 10  & 0.01 -- 0.3 &  0.0 -- 49.0  & $6\times6$ &$V^{0}$\\
    4 & 100 & 0.01 -- 1.0 &  0.0 -- 40.0  & $1\times1$ &$V^{0}$\\
    5 & 100 & 0.01 -- 3.0 &  50.0         & $1\times1$ &$V^{0}$\\
    6 & 100 & 0.01 -- 1.0 &  50.0         & $1\times1$ &$V^{0}$\\
    \hline
    A & 300 & 10 & 0, 4 &  $1\times1$ & $V^{0} V^{01} V^{10}$ \\
    B & $3000$ & 10 & 0, 1, 2, 3, 4 &  $1\times1$ & $V^{0}$\\
    C & $3000$ & 10 & 0, 1, 2, 3, 4 &  $1\times1$ & $V^{0}$\\
    \hline
  \end{tabular*}
  \label{tab:ptc}
\end{table}

\subsection{Uniformity measurements}
\label{sec:uniformity}

The measurement of the uniformity of the camera response is similar to
that of the stability, but the camera is moved through 25 positions in
a $5\times5$ grid spanning a square of 60mm side with 15mm resolution
in the flat-field beam. The goal is to disentangle variations in the
camera response from spatial variations of the illumination. The
measurement is only carried at a sensor temperature of
\SI{0}{\celsius} but is repeated for the sixteen LEDs to look for
wavelength dependencies. All LEDs are operated at a stabilized current
of \SI{10}{mA} where they deliver of the order of \SI{60}{ke^-/s} in a
single camera pixel. The exposure time is set to \SI{0.3}{s} so that
none of the pixels saturate.

\subsection{Quantum efficiency measurements}
\label{sec:qemeasuremen}

The quantum efficiency measurement is performed in the range
$375$--\SI{1078}{\nano\metre}, turning on each of the 26 relevant
LEDs, one at a time. After turning on one LED, the camera is set to
intercept the monochromatic beam, and the corresponding wavelength
range is scanned with a \SI{1}{\nano\metre} step size. For each
wavelength, two successive camera exposures are taken, one with the
beam shutter (S) closed and a second with the beam shutter
opened. Simultaneously the same exposures are obtained on the
monitoring sensor. Once the wavelength range is exhausted, the
reference photodiode is brought to intercept the monochromatic beam,
and a similar scan is performed using the monitoring sensor and the
photodiode. The quantum efficiency estimate is built from these 4
measurements as:
\begin{equation}
  \label{eq:qe}
  QE(\lambda) = \frac{\Phi G}{\tau_{exp}} \frac{e\epsilon_{NIST}(\lambda)}{I_{NIST}}\frac{M_N}{M_C}
\end{equation}
where $\Phi$ is the measured ADU count in the camera, $G$ is the
camera gain estimate in \si{e^-/ADU}, $\tau_{exp}$ is the exposure
time in seconds, $\epsilon_{NIST}$ is the quantum efficiency of the
reference sensor in \si{e^-/\gamma}, $I_{NIST}$ is the photocurrent
delivered by the reference sensor in ampere, $e$ is the elementary
charge in coulomb, and $\frac{M_N}{M_C}$ is the ratio of the count
rate in the monitoring sensor during the reference photodiode and
camera exposures. We further detail the computation of the different
terms in what follows.

\subsubsection{Photodiode operations}
\label{sec:photodiode}

\begin{figure}
  \centering
  \includegraphics{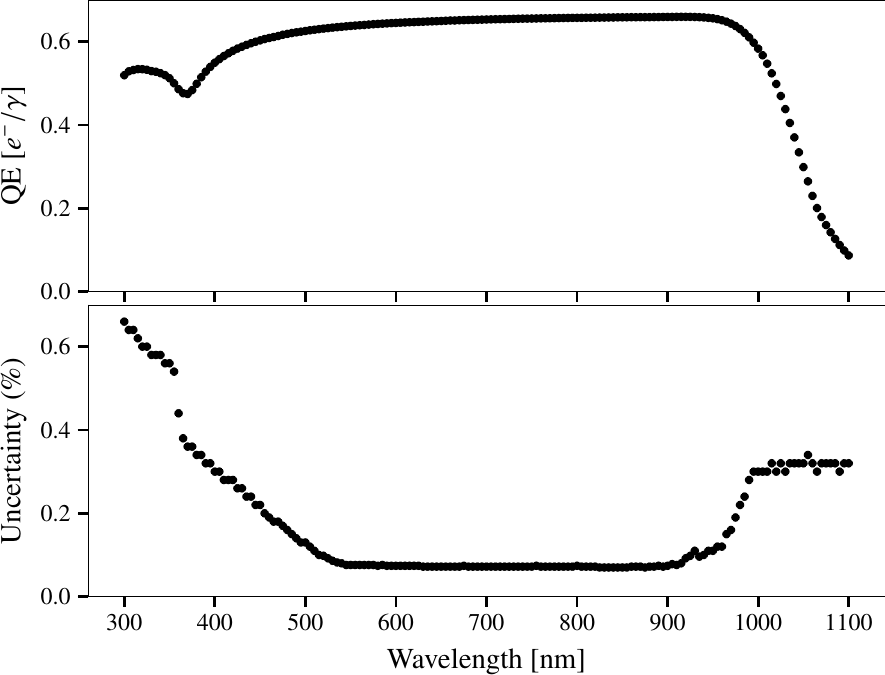}
  \caption{\emph{Top:} Quantum efficiency curve of our
      reference Hamamatsu S2281 photodiode as measured by the
      NIST. \emph{Bottom:} Quoted relative uncertainty on the quantum
      uncertainty at the \SI{95}{\percent} level ($2\sigma$).}
  \label{fig:NIST}
\end{figure}

Our reference sensor is a Hamamatsu S2281 photodiode, with a circular
photosensitive area of \SI{1}{\cm\squared} (\SI{5.65}{\mm} radius). The measurement
of its quantum efficiency curve $\epsilon_{NIST}(\lambda)$ performed at NIST
\citep{houston2008detectors} along with associated uncertainties is
recalled in Figure~\ref{fig:NIST}.

Originally built with a non-cooled camera in mind the heat generated
by the focal plane cooling slightly exceeded the heat dissipation
capability of the sensor calibration bench enclosure resulting in a
nominal operation temperature in the range
\SIrange{26.4}{26.6}{\celsius}. This temperature is slightly larger
than the target \SI{23}{\celsius} temperature at which the reference
NIST sensor was calibrated. This introduces a systematic uncertainty
in the infrared region due to the change of the photodiode quantum
efficiency with temperature. According to \citet[Fig
9.1]{houston2008detectors} the change is only important beyond
\SI{1000}{\nm} where it amounts to \SI{0.2}{\% / \celsius}.

Centering of the photodiode in the monochromatic
light beam is performed by scanning in x and y at constant illumination
past the edges of the photodiode sensitive area. The photodiode is
then moved back to the barycenter of the resulting flux map.

\begin{figure}
  \centering
  \includegraphics{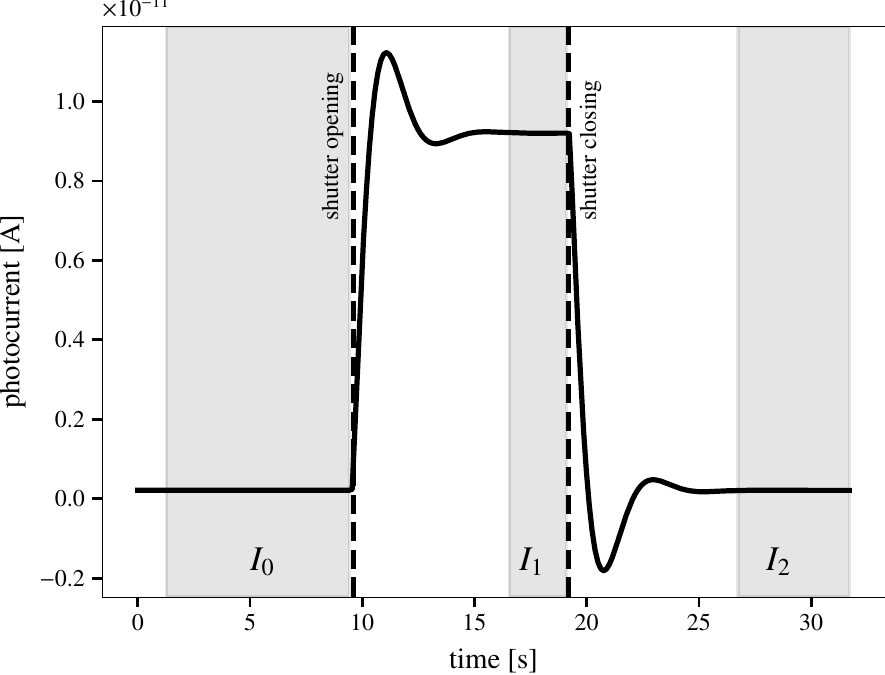}
  \caption{Average shape of the photocurrent reading in the
    NIST-calibrated sensor. The shaded regions corresponds to readings
    selected to build the photocurrent estimate.}
  \label{fig:photocurrent}
\end{figure}

The flux level of the monochromatic beam was designed to be relatively
low in order to avoid saturation of CCD cameras in relatively long
exposures (larger than \SI{1}{\s}), allowing its use to calibrate
cameras with mechanical shutters while minimizing exposure time
uncertainties. As a consequence, sensitive readings of the
photocurrent required us to select a large feedback resistance in the
transimpedance amplifier and limited bandwidth.  The photocurrent out
of the photodiode is sampled at \SI{7.8}{\hertz} by a Keithley 6514
picoammeter. At any given wavelength the photocurrent is sampled for
\SI{30}{\s}, starting \SI{10}{s} before the opening of the
monochromatic beam shutter. The shutter is left open for \SI{10}{\s}
and closed again for the end of the operation. The average shape of
the photocurrent reading is plotted in Fig.~\ref{fig:photocurrent}. We
construct 3 numerical averages $I_0$, $I_1$ and $I_2$ of the readings
before, during and after the light exposure, avoiding the transition
regions. We construct the photocurrent estimate as
$I_{NIST} = I_1 - (I_0 + I_2)/2$, and estimate uncertainties on
$I_{NIST}$ from the standard deviation of the samples in each of the
quantities as:
$\sigma_{I_{NIST}}^2 = \sigma_{I_1}^2 + \frac{\sigma_{I_0}^2 +
  \sigma_{I_2}^2}{4}$.

\subsubsection{Camera operations during quantum efficiency
  measurements}

We apply the same settings for the quantum efficiency measurement as
for the PTC measurement, so that we can assume the readout gain to be
identical. In normal mode, we select $2s$ exposures for both the
``dark'' and ``open'' exposures.

The estimate of the camera count $\Phi$ at each wavelength is built in
several steps. First we compute the centroid of the slit image in the
camera by building a stack of all images from a complete wavelength
scan. The stack is built as the sum of all open images minus the sum
of all dark images. Second, we perform the photometry of all images in
circular apertures centered on the computed centroid with physical
radii of $2$, $5.65$ and \SI{6}{\mm}.\footnote{Converted to pixels
  assuming a pixel size of \SI{3.76}{\micro\metre}.} The second radius
matches exactly the physical size of the NIST photodiode, and is very
significantly larger than the slit image FWHM of
\SI{2.9}{\milli\metre}. This aperture minimizes systematic errors from
small amount of light diffused at large angles (see the discussion in
Sect.~\ref{sec:qeresults}). The flux in an aperture of radius $r$ is
denoted $\Phi_r$. We also gather the sum of all pixels in the image
$T$ and compute a background light proxy as $B = T - \Phi_6$.

\begin{figure}
  \centering
  \includegraphics{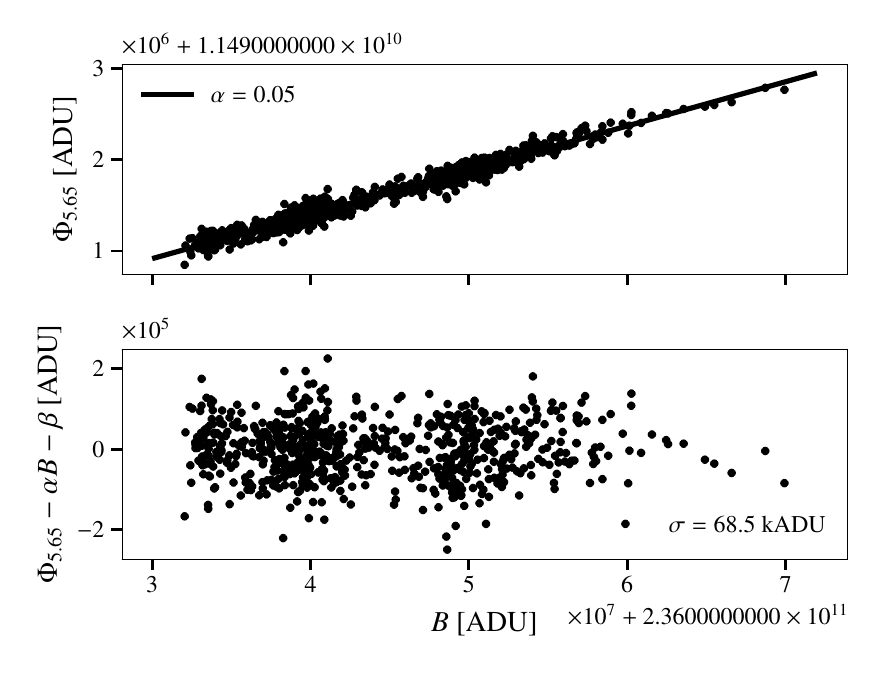}
  \caption{\emph{Top:} Instrumental flux in \SI{5.65}{\mm} apertures
    measured in ``dark'' images as a function of the background light
    proxy build from the never exposed region of the sensor. The
    best-fit linear relation (solid black line) gives an estimate of
    background contribution in exposed images. \emph{Bottom:}
    Residuals to the best fit relation. The measured rms gives the
    level of the background subtraction noise which is the dominant
    noise contribution in our QE measurement.}
  \label{fig:background}
\end{figure}

The large apertures, required to match the physical size of the
reference sensor, makes the photometry very sensitive to background
contamination. We use the ``dark'' exposures to refine our background
estimate as follows. For a given aperture $r$, we adjust a linear
relation between $B$ and $\Phi_r$ on all the ``dark'' images in the
sequence as $\Phi_r = \alpha B + \beta$. An illustration of the fit is
given in Fig.~\ref{fig:background}.
We then use the best fit parameters to build the background corrected
aperture flux as $\bar \Phi_r = \Phi_r - \alpha B - \beta$. Doing so,
the uncertainties on background subtraction are typically of the order
of \SI{70}{\kilo ADU} for the full \SI{5.65}{\mm} aperture and drops
down to \SI{15}{\kilo ADU} for the smaller \SI{2}{\mm} aperture where
it compares more favorably to the photon noise (see the results
section for a more detailed discussion of the various noise
contributions).

\subsubsection{Monitoring sensor}
\label{sec:monitoring}

Lastly, we use an IMX174 sensor to monitor changes in illumination
while switching between target and reference sensor. When the camera
faces the beam, the exposures of the monitoring sensor are
synchronized with the camera exposures and the resulting photometry
delivers the quantity $M_c$ in Eq.~(\ref{eq:qe}). With the photodiode
in the beam, the monitoring sensor continuously acquires \SI{2}{\s}
exposures. Images corresponding to opening and closing of the shutter
are discarded, and the photometry of the three fully exposed images is
averaged to deliver the quantity $M_N$ in Eq.~(\ref{eq:qe}).

In both case, photometry is performed similar to what was described
for the target sensor: we gather photometry in several apertures
centered on the centroid of the slit images, and build a background
estimator from ``dark'' images. The only difference is that we are not
constrained to match the aperture size of the NIST photodiode for this
monitoring measurement, the only important point being that $M_N$ and
$M_c$ are derived from the same aperture so that the ratio of the two
corresponds to the variation of the illumination. Therefore we select
a much smaller aperture (\SI{1}{mm}) closer to the extent of the
image, to optimize the signal to noise ratio.

\section{Results}
\label{sec:results}

\subsection{Sensor cosmetics and inter-pixel capacitance}
\label{sec:ipc}

The sensor cosmetics is studied on a stack of 3000 dark frames with
10s exposures with a stabilized sensor temperature of
\SI{0}{\celsius}. The fractions of pixels displaying a level of dark
current in excess of \num{1}, \num{10}, \num{100} and \SI{1000}{e^- / \s} are
respectively \num{5.00e-4}, \num{3.69e-5},
\num{4.11e-6} and \num{2.08e-7}.

\begin{figure}
  \centering
  \includegraphics{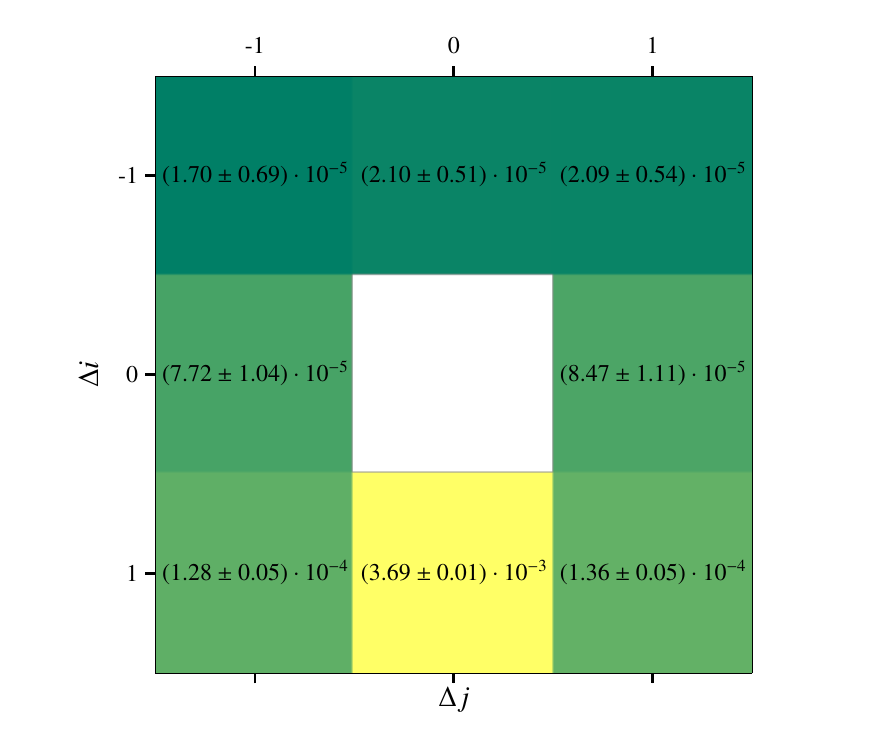}
  \caption{Coupling coefficients between neighboring pixels
    attributed to interpixel capacitance (IPC). The asymmetry of the
    effect imposes to distinguish between even and odd column and line
    numbers. In the figure the pattern is presented as it would appear
    for a central pixel with even line and column coordinates. A pixel
    with odd line coordinates is instead mostly correlated with the
    pixel immediately above.}
  \label{fig:ipc}
\end{figure}

We can take advantage of the hot pixels to get a handle on the pattern
of the inter-pixel capacitance
(IPC, \citealt{2004SPIE.5167..204M,2006ASSL..336..477F}), which is found to
be significant only between two adjacent pixels in a column, with a
correlation coefficient of $3.69 \pm 0.01 \cdot 10^{-3}$. The effect
is not symmetric. Pixels with even line numbers are only coupled to the
pixel in the following line (respectively odd-parity pixels are
coupled to the preceding pixels). The effect on photon statistics is
thus half what the correlation constant would suggest. The measured
$3\times3$ correlation pattern between neighboring pixels as seen by
pixels with even-parity coordinates is presented in
Fig.~\ref{fig:ipc}.

\subsection{Stability of the photometric response}
\label{sec:stabilityresults}

\begin{figure}
  \centering
  \includegraphics{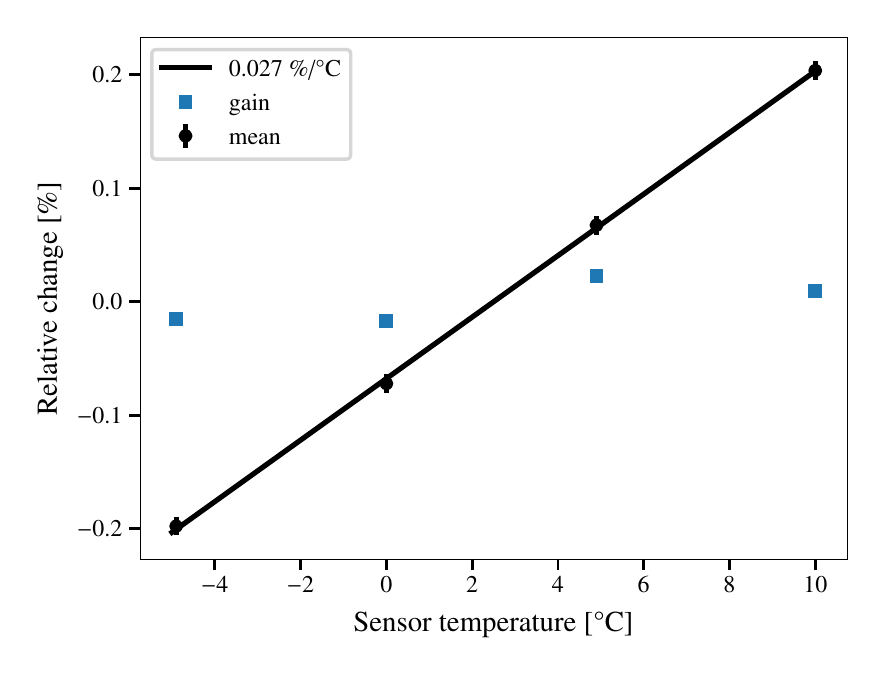}
  \caption{Relative change in camera response (black circle) and gain
    (blue squares) with sensor temperature. The black line is a linear
    fit to the change in camera response.}
  \label{fig:temperaturetrend}
\end{figure}

The stability of the camera response in time and temperature is
obtained from the stability dataset described in
Sect.~\ref{sec:stability}.  A small trend with sensor temperature is
detectable in the camera response. Fig.~\ref{fig:temperaturetrend}
displays the measured response at 4 different stabilized temperatures
between $-5$ and \SI{10}{\celsius}. The measurements are well
described by a simple linear trend, with a relative slope of
\SI{0.027}{\% / \celsius}. The gain estimate derived from the same
dataset is in contrast extremely stable, and compatible with no gain
change at the $2\cdot10^{-4}$ level over the \SI{15}{\celsius} spanned
by the test data. This result suggests that the observed temperature
trend is dominated by a change in the sensor quantum efficiency. The
test, conceived as a check of the gain stability, was performed at a
single wavelength. It would be interesting to repeat the same
measurement at different wavelength, especially in the infrared
where the strongest temperature dependence is expected.

The result shows that sub-mmag photometric stability is easily
achievable even without stringent control of the environment, in
particular when operating at a steady framerate. Here, the RMS of the
camera average in the second batches of stability exposures (that is,
the batch starting after 100 stabilization exposures) is
\num{3.26e-5} on average. This is even lower than the RMS of the
monitoring photodiode readings ($\sim10^{-4}$ see
Fig.~\ref{fig:illuminationstability}), so that we cannot distinguish
between fluctuations in the camera response or in the
illumination. The quoted number can be used as an upper bound on the
camera response fluctuation in stable conditions.

\subsection{Study of the readout electronics}
\label{sec:PTCFIT}

The shape of the relation between the variance and the expectation of
the pixel readout across a large range of illumination levels, usually
called the photon transfer curve, is a well-documented
tool to study the gain of the readout chain of CCD sensors. It was
recently extended to the study of the dynamical electrostatic effect,
coined as brighter-fatter effect, affecting thick deep-depleted CCD
sensors (see \citealt{astier} for the modification of the PTC shape
and reference therein for a description of the effect).

As a large number of pixels share the exact same readout chain, CCD
studies typically rely on statistics built from a large number of
pixels and a small number of images.\footnote{Differences in pair of
  images are typically used to suppress the effect of structure in the
  illumination pattern.} Different levels of spatial averaging are
relevant in the study of CIS: averaging over the entire focal plane
gives the best handle over features shared by the entire ADC array,
such as reference voltages, averaging over individual columns is
expected to show defects specific to a single ADC pipeline. Measuring
the gain for individual pixels through this technique is however
challenging and requires very large statistics. Around \num{30000}
images are required to reach percent accuracy on the individual pixel
gain. We present limited but promising efforts in this direction in
the next section. As no information is available to us regarding the
structure of the sensor readout chain we only assume that it follows a
standard column parallel structure.

We propose various models for the statistics $M_p$ and $V^0_p$ and
$m_i$ in appendix \ref{sec:stat-model-phot}. In the case where perfect
linearity of the pixel response to illumination can be assumed, and
where the photo-conversion is a perfect Poisson process (independent
conversion events with constant probability), the readout chain is
described by 3 parameters: its gain $G$ in \si{e^- / ADU}, the readout
noise $\sigma$ and the readout bias $b$, which can be inferred from
fitting relation~(\ref{eq:result}) to the empirical
statistics. Deviations from the linearity or Poissonian hypothesis
will appear as deviations from the straight line in the
plot. Disentangling between the different effects requires additional
data, either direct linearity measurements or extensive study of the
correlation between neighboring pixels.

A simple step to mitigate the effect of correlations between
neighboring pixels (either from IPC or brighter-fatter) is to compute
statistics on pixels made artificially bigger by binning. We first
present the study of the average transfer curve of the sensor,
obtained by binning pixel counts in $6\times6$ square superpixels, on the
first third of the digital scale which widely encompasses the dynamic
range over which the quantum efficiency measurements has been
performed. We then present the measurements of the PTC on non-binned
statistics over the full-scale as well as direct check of the integral
linearity, and nearest neighbors covariances. Those three measurements
combined hints toward a consistent picture with good integral
linearity and measurable deviation from the Poisson hypothesis. At
this stage, however, we cannot provide a complete model satisfactorily
describing this dataset. Therefore, we rely on the measurement in
binned pixels to infer our baseline value for the camera
gain. Finally, we present our measurements of the readout gain
uniformity.

\subsubsection{Photon transfer curve in $6\times6$ superpixels}
\label{sec:DNL}

\begin{figure*}
  \centering
  \includegraphics{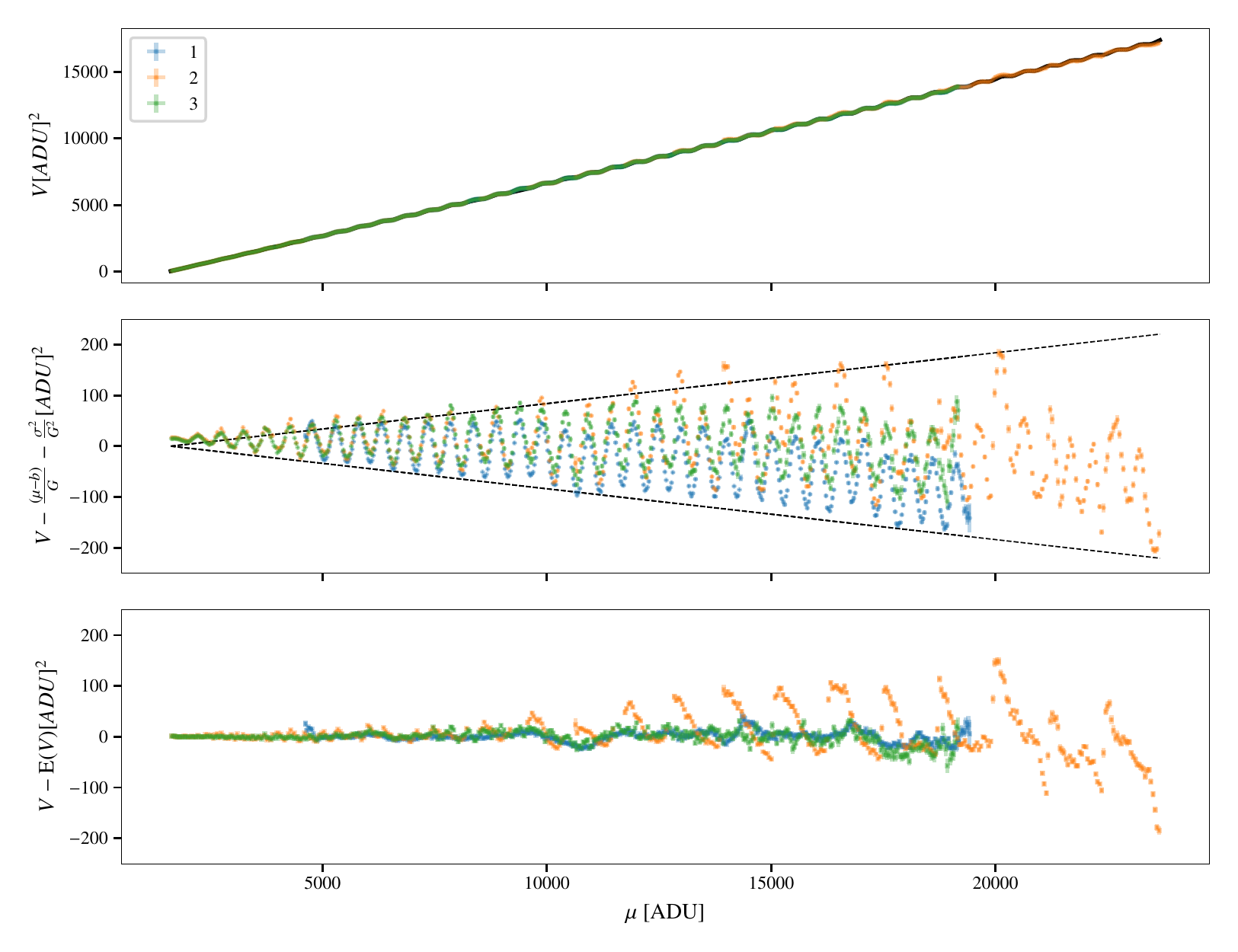}
  \caption{\emph{Top:} Photon transfer curve for the IMX411 sensor
    obtained from statistics build from $6\times6$ \emph{binned
      superpixels} over the first third of the scale. The 3 colors
    corresponds to 3 independent datasets obtained with different
    illumination, exposure time and number of images (see
    Table~\ref{tab:ptc}). The black solid line presents the linear
    relation from Eq.~(\ref{eq:result}) using the best fit gain
    value. \emph{Middle}: Residuals to the linear model. The dashed
    black lines figure 1\% variations of the integral
    gain. \emph{Bottom}: Residuals to the full model, including the
    effect of a 512-bit periodical DNL with 0.65 ADU of amplitude, and
    provision for curvature of the PTC due to residual influence of
    brighter-fatter effect in the binned superpixels.}
  \label{fig:ptc}
\end{figure*}

For each $M_p$ and $V_p$ in datasets 1--3, we compute the mean
variance $V$ of all the $N_\mu$ superpixels sharing the same
$M_p = \mu$ value as follows:
\begin{equation}
  \label{eq:binnedvar}
  V(\mu) = \frac{1}{N_\mu}\sum_{\lbrace p, M_p=\mu \rbrace} V_p - \Delta^2(\mu-b)^2\,, 
\end{equation}
where the $\Delta$ term corrects the statistics for the effect of
small illumination fluctuations during the measurement (see
appendix~\ref{sec:variation} for details). This procedure allows us to
recover the full resolution of the PTC shape, while a simple averaging
over the focal plane would smear the features due to non uniformity in
the illumination level. The result is presented in
Figure~\ref{fig:ptc}. All three datasets are presented on the same
plot, and present a rather consistent picture.

The most striking feature is a periodic differential non linearity
(DNL) very consistent in all datasets, likely corresponding to a fix
bit width error. As striking as it appears on the PTC the DNL likely
has little practical consequences. However, it complicates the
analysis of the PTC as statistics obtained with slightly different
illumination are no longer directly comparable and a simple model for
the shape does not describe the data.

When fitting independently a simple linear relation to the three
datasets, we recover an average gain value of
\SI[separate-uncertainty]{1.274 \pm 0.0016}{e^- /ADU}, where the
quoted uncertainty is the RMS between the three datasets. The middle
panel in Figure~\ref{fig:ptc} shows the residuals to this simple model
for the three datasets. Fitting instead for the approximate shape of
the PTC in presence of BF proposed in \cite[Eq. 16]{astier}, gives
a significantly different average value of
\SI[separate-uncertainty]{1.2636 \pm 0.0023}{e^-/ADU}, a $4-\sigma$
difference, with a moderate preference for a non zero curvature of the
binned PTC: $a_{00} = -5 \pm 2 \cdot 10^{-7}$. The recovered values
for the curvature are not very consistent between the three datasets,
and they change when the top of the scale is cut differently (e.g. at
\SI{15}{\kilo ADU}) as can be expected from the poor quality of the
fit.

A rough model of the DNL can be attempted to improve the fit
consistency and enable comparison between statistics at different
focal plane locations more easily, alleviating the need for matched
illumination levels.  In appendix \ref{sec:variablebit}, we present
the tools to numerically compute the relevant statistics with
arbitrary digital boundaries. We model the DNL as a sine function with
a period of 512 bits, that is to say, we assume that the digital scale
admits boundaries between codes
$D_n = n + A_{DNL}\cos\left(\frac{2\pi}{512}
  (n-\phi_{DNL})\right)$. Not knowing the details of the architecture
of the ADC array, little more than that can be done, but this ad-hoc
model may be a good-enough description of the reality for the need of
the present study. Fitting for $A_{DNL}$ and $\phi_{DNL}$ along with
the readout gain and noise delivers a fairly comprehensive description
of datasets 1 and 3, although distinctive features remain visible in
the residuals as shown in the bottom panel of Fig.~\ref{fig:ptc}. The
best fit values are $A_{DNL}= 0.651\pm0.006 \si{ADU}$ and
$\phi_{DNL}=44.4\pm0.2$.  The corresponding reduced chi-squared for
the two datasets are $1.41$ and $1.22$. Interestingly, dataset 2 which
is obtained from a much higher number of consecutive frames (see
table~\ref{tab:ptc}) is at odds with the other two datasets with way
less satisfactory fit quality of $\chi^2/{d.o.f.} = 2.39$. The reasons
behind this disagreement are for the moment not understood. However
the problem is unlikely to affect the gain determination as all three
datasets now deliver a very consistent gain value
$G = 1.273 \pm 0.0008 \si{e^-/ADU}$, where again the quoted
uncertainty is the RMS of the 3 fits. The improved modeling, however,
does not entirely solve the inconsistency between the datasets when
allowing for curvature. The mean and RMS of the recovered curvature
now settles at $a_{00} = -1.6\pm 1.3 \cdot 10^{-7}$, compatible with
zero and resulting in a slightly lower mean gain value of
$G = 1.269 \pm 0.0035 \si{e^-/ADU}$. We adopt this last value and its
uncertainty as our baseline determination of the readout gain. Due to
the binning, this value is not affected by the IPC.

\subsubsection{Distinguishing between non linearity and other effects in the full scale PTC}
\label{sec:INL}

\begin{figure*}
  \centering
  \includegraphics{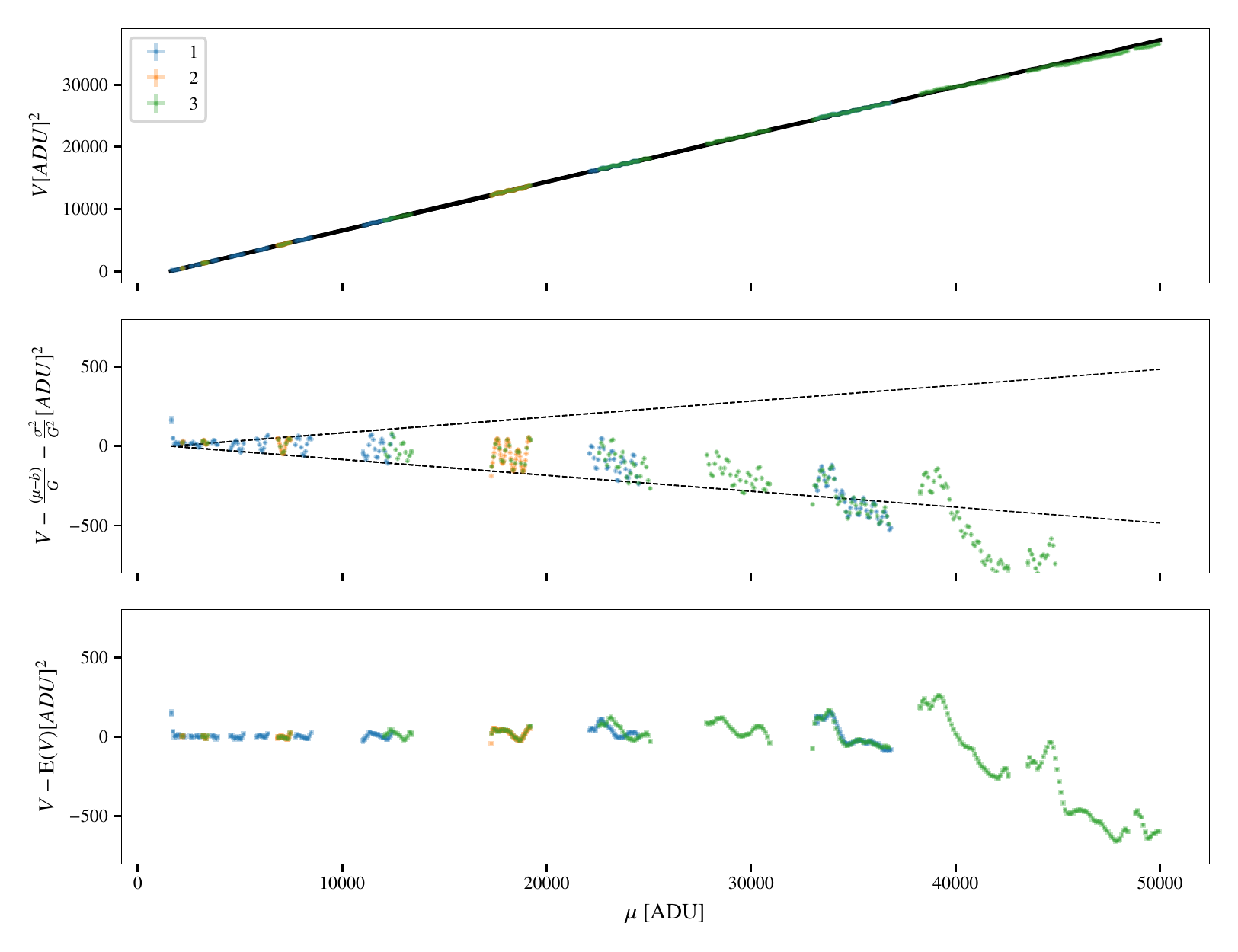}
  \caption{\emph{Top:} Photon transfer curve for the IMX411 sensor
    obtained from statistics build from \emph{individual physical
      pixels} over the full scale. The 3 colors corresponds to 3
    independent datasets obtained with different illumination level
    and exposure times (see Table~\ref{tab:ptc}). The
    black solid line presents the linear relation from
    Eq.~(\ref{eq:result}) using the best fit gain
    value. \emph{Middle}: Residuals to the linear model. The dashed
    black lines figure 1\% variations of the integral
    gain. \emph{Bottom}: Residuals to the full model, including the
    effect of a 512-bit periodical DNL with 0.65 ADU of amplitude, and
    provision for curvature of the PTC due to influence of
    brighter-fatter-like effect in the physical pixels.}
  \label{fig:ptcfull}
\end{figure*}

The PTC measured directly on individual pixels and over the full scale
is presented in Figure~\ref{fig:ptcfull}. The model including both
brighter-fatter and 512-bit periodic non-linearity is a rather poor
fit of the data, mainly because other differential features show up
beyond the 512 bit-periodic feature. The best-fit parameters are
however fairly compatible with those previously obtained:
$G=\SI{1.273 \pm 0.007}{e^-/ADU}$, $A_{DNL}= \SI{0.665\pm0.016}{ADU}$, and
$\phi_{DNL}=49.2\pm1.7$. The uncertainty on the PTC curvature remains
high: $a_{00} = -4.0 \pm 1.9 \cdot 10^{-7}$, because of the remaining
features. We have not attempted to build a comprehensive model of the
remaining features at this stage, but reckon that the general shape of
the PTC points toward significant non null curvature. Correcting the
recovered gain value from the expected influence of the IPC measured
on hot pixels (see sect~\ref{sec:ipc}), brings the gain estimate on
the full statistics even closer to the value recovered in binned
statistics.

\begin{figure}
  \centering
  \includegraphics{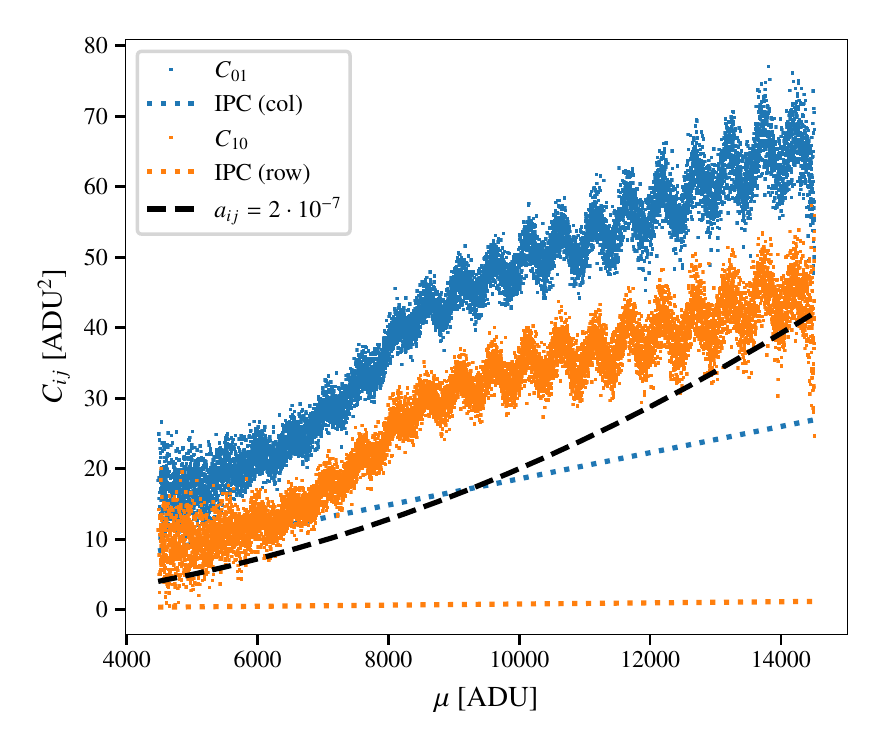}
  \caption{Measured covariances between the nearest neighbors in a
    line ($C_{01}$) and in a column ($C_{10}$) as a function of pixel
    readout. The dashed lines figure the corresponding expected
    covariances from the measured IPC. The black dashed line figures
    the first-order shape of the correlation expected for an amplitude
    of the correlation factor corresponding to $\frac{-a_{00}}{2}$,
    with $a_{00}$ the best fit value from the PTC.}
  \label{fig:c0110}
\end{figure}

\begin{figure}
  \centering
  \includegraphics{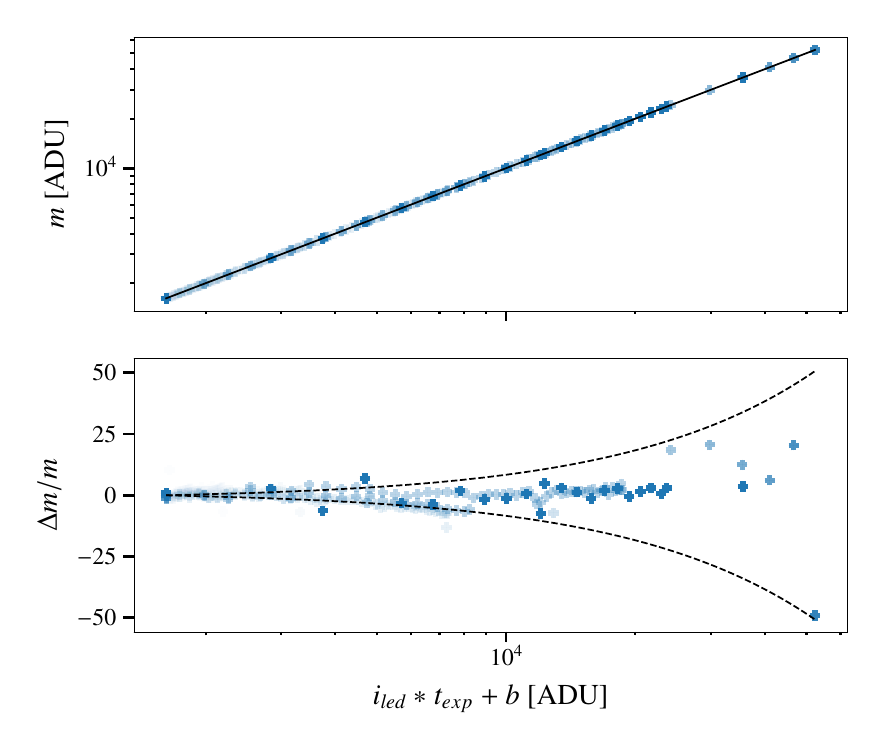}
  \caption{Direct linearity check performed by interlacing
    measurements at increasing exposure time and increasing
    brightness. Dot intensity encodes the exposure time. \emph{Top:}
    Average illumination level in the camera as a function of the
    modeled illumination intensity in the flat-field
    source. \emph{Bottom:} Difference between measured camera level
    and the modeled flat-field source intensity. The dashed black
    lines figure variation of the integral gain by $\pm0.1\%$.}
  \label{fig:linearity}
\end{figure}

Both classical non linearity and brighter-fatter effect could cause
the observed curvature of the PTC. Behavior similar to the
brighter-fatter has been previously reported in
CMOS \citep{2021arXiv211201691G} and CMOS NIR
detectors \citep{2018PASP..130f5004P}, with a proposed physical
mechanism, shrinkage of the depletion
region \citep{2017JInst..12C4009P}, paralleling the mechanism observed
in CCDs. Disentangling between classical non linearity or
brighter-fatter effect necessitates either direct linearity
measurements or detection of the characteristic correlation between
neighboring pixels introduced by the electrostatic influence. We
present both measurements in Figures~\ref{fig:c0110} and
\ref{fig:linearity}.

The dataset specifically taken to study correlations (A in
Table~\ref{tab:ptc}) provides moderate evidence for the presence of
positive correlations between the nearest neighbors in a line
($C_{01}$) and in a column ($C_{10}$), in excess of what would be
expected from measured IPC alone. The resulting curves are shown in
Figure~\ref{fig:c0110}. The shape does not readily comply with
expectations for BF-induced correlations and we don't have at this
stage a compelling model for the measured shape. The most obvious
caveat in the data-set is the need to subtract for the contribution of
quite large illumination fluctuations ($\Delta \sim 10^{-3}$). This is
done using the RMS of the mean of all images in the sequence to build
an estimate of the illumination fluctuation (see
appendix~\ref{sec:variation}). We note that the proposed technique
would allow investigation of the correlations at larger distances
$C_{ij}$, however adequate hardware to perform the required
comparatively heavy computations in real time would be required.
The flat-field illumination source was built with the idea of
providing a direct linearity measurement, but at this stage the
linearity of the illumination has not yet been calibrated. The
gathered dataset still allows limited investigation of the linearity
by interlacing variations of the illumination (through variation of
the LED intensity), with variations of the exposure time, and solving
for illumination level at a given LED intensity. Doing so gives the
linearity check presented in Fig.~\ref{fig:linearity}, which does not
detect significant integral non-linearities in the chain up to
\SI{55000}{ADU} (\SI{70}{\kilo e^-}).

We therefore conclude that the curvature observed in the PTC is not
attributable to classical non-linearity and is at least qualitatively
compatible with expectations from a small brighter-fatter effect.

\subsubsection{Uniformity of the readout gain}
\label{sec:adc}

\begin{figure*}
  \centering
  \includegraphics{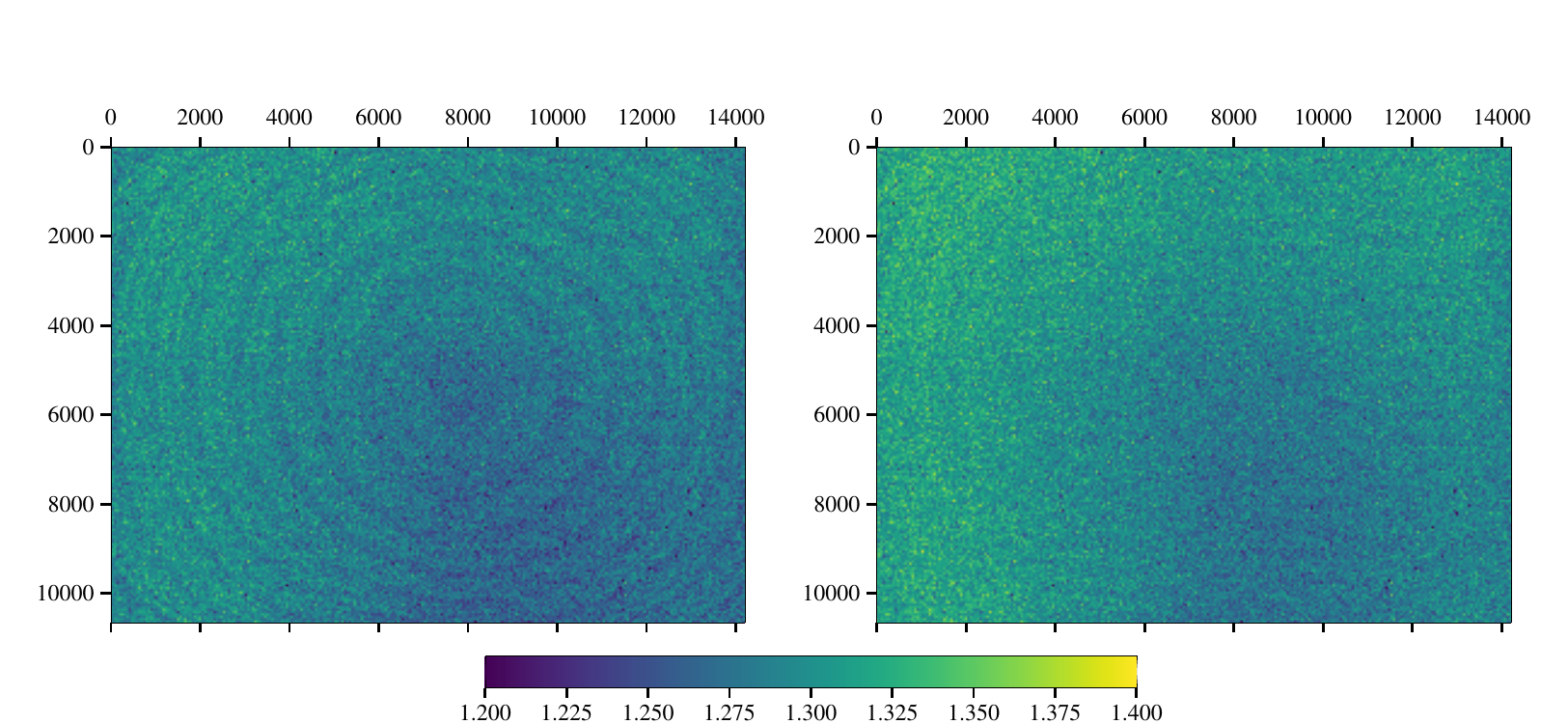}
  \caption{\emph{Left:} Uncorrected map of the individual pixel gain
    obtained from 12000 images at 4 different illumination levels. The
    circular features arise from a periodic differential non linearity
    imprinted on the gain estimate due to a center-to-edge gradient in
    the illumination pattern. \emph{Right:} Same as before but
    correcting the measured empirical variance for the prediction of
    the non-linearity model adjusted in Fig.~\ref{fig:ptc}.}
  \label{fig:gainmap}
\end{figure*}

Determination of the individual pixel gains using this technique
requires a very large number of images, but is in principle
achievable. We made an attempt to test the concept with the high
statistics datasets denoted (B) and (C) in Table~\ref{tab:ptc}. The
result is presented in the left panel in Fig.~\ref{fig:gainmap}. The
main visible feature in the recovered gain map is the imprint of the
differential non-linearity due to the smooth change in illumination
level in the flat-field images. Simply correcting the measured
variance using the ad-hoc model from Fig.~\ref{fig:ptc}, gives the
per-pixel mean-over-variance map presented on the
right-panel. Residual features in the corrected map show that the
applied correction is not fully adequate across the focal plane, but
at least qualitatively describe the phenomenon.  The corrected map has
an RMS of $1.7\%$, dominated by statistical noise. Binning the map by
$120\times120$ pixel to suppress the statistical noise brings the RMS
down to $1.0\%$, with residual artifacts from the DNL dominating the
spatial features. Therefore we conclude that we can set an upper bound
of $1\%$RMS on spatial variations of the effective gain value across
the focal plane. In addition the average gain in the region
illuminated in the quantum efficiency measurement differs from the
focal-plane average by \SI{0.0008}{e^-/ADU}. Therefore, we stick to
the baseline gain value to interpret our quantum efficiency
measurement.

\subsection{Uniformity of the camera response}
\label{sec:uniformityresults}

The camera response (\emph{i.e.} the product of the quantum efficiency
and the readout gain) has been mapped at the 16 wavelengths delivered
by the flat-field light source. Given that the spatial extent of the
IMX411 sensor is larger than the area over which the flat-field beam
can be considered uniform, we resorted to iteratively solve for the
intensity of the illumination pattern and the camera response map. The
25 images in a single wavelength dataset are thus interpreted as the
product of the illumination pattern $B(\vec x)$ at this wavelength
with the camera response map $M(\vec x)$ for 25 known central
positions $x_i$ of the camera. Image $i$ therefore follows the
equation:
\begin{equation}
  \label{eq:conv}
   I_i(\vec x) = B(\vec x + \vec x_i) \times M(\vec x)\,.
\end{equation}
The illumination pattern over the full area is developed on a cardinal
bicubic B-spline basis with $10\times10$ nodes. A first estimate of
the illumination pattern $B^0$ is obtained assuming a uniform camera
response $M^0=1$, and this $B^0$ is used to solve for a first map of
the camera response $M^1$. The procedure is iterated once and quickly
converges. The model is an accurate description of the dataset except
for projective features such as defects on the camera window.

\begin{figure*}
  \centering
  \includegraphics{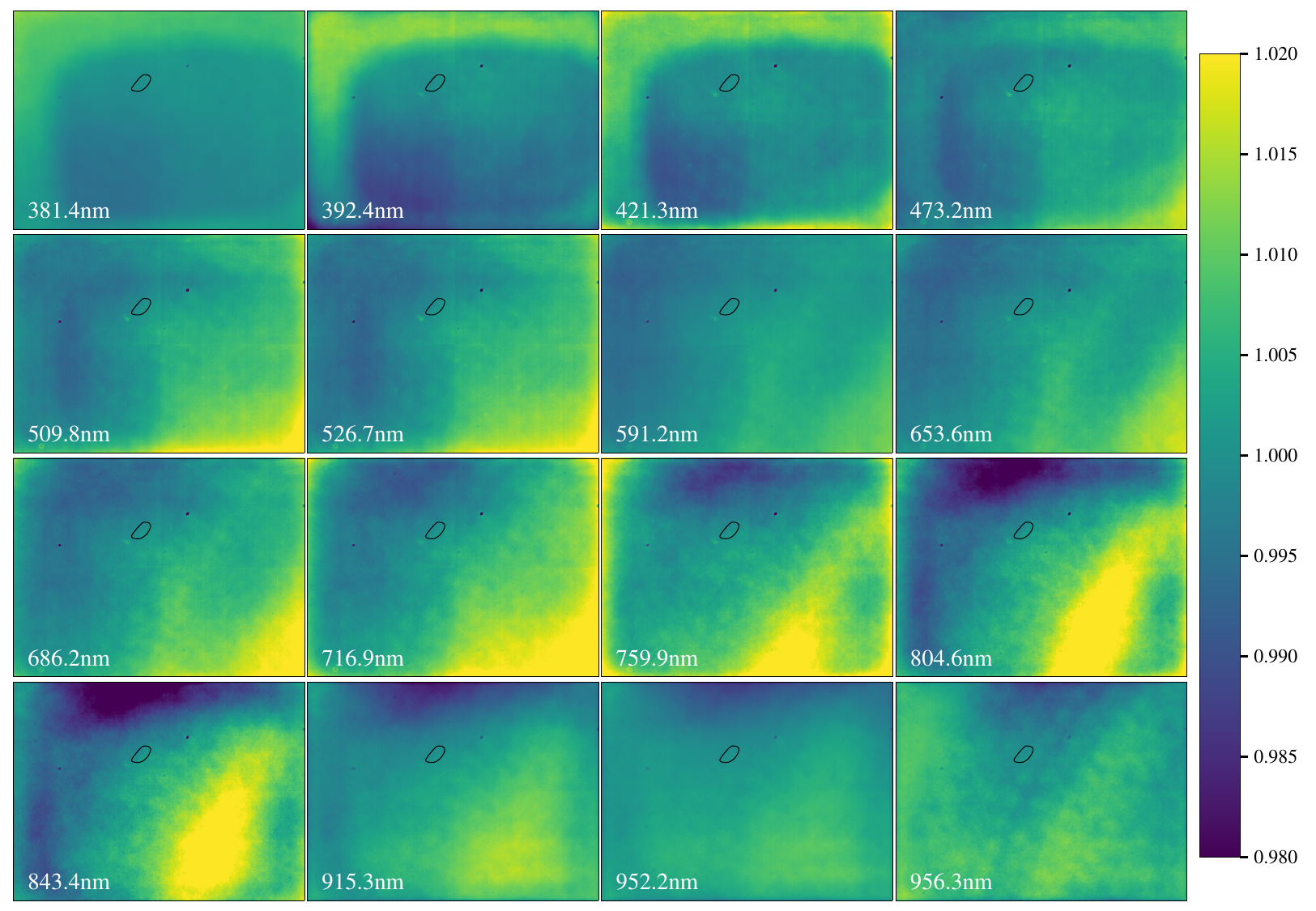}
  \caption{Map of the transmission uniformity for the 16 different
    flat-field channels, corresponding to different illumination
    wavelength from UV to near infrared. The scale of the map is
    relative to the average response over the entire sensor area. The
    black contour figures the region in which the high-resolution
    quantum efficiency curve measurement has been performed.}
  \label{fig:qemap}
\end{figure*}
The resulting camera response maps are shown in
Figure~\ref{fig:qemap}. All demonstrate uniformity in the response at
the level of $4\%$ peak-to-peak ($<1\%$ RMS), with the position
selected for the QE measurement being in general slightly below the
average. The pattern shows some evolution with wavelength, however,
the evolution appears slow enough to be corrected by filter-dependent
flat-fielding, with an accuracy dependent on the chosen photometric
system.

\subsection{Quantum efficiency curve}
\label{sec:qeresults}

\begin{figure*}
  \centering
  \includegraphics{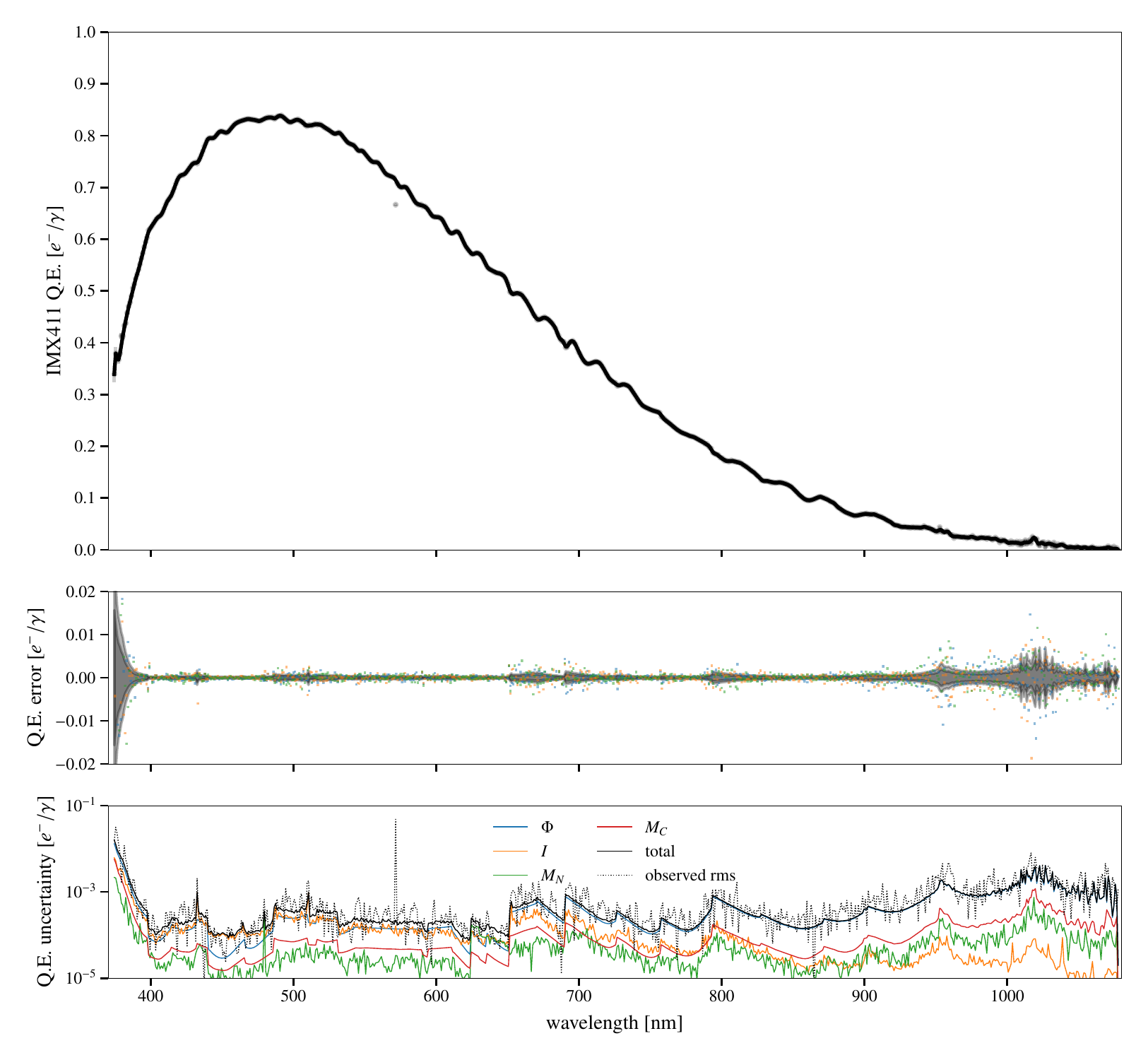}
  \caption{\emph{Top:} Quantum efficiency curve of the QHY411M camera
    (window included). Three independent measurements at the same
    location are superimposed. The solid black line correspond to a
    smooth B-spline fit to the data. \emph{Middle:} Residuals to the
    B-spline fit. The three colors correspond to the three independent
    measurements. \emph{Bottom:} Breakout of the noise in the quantum
    efficiency measurement according to the origin of the contribution
    from the four different terms in Eq.~(\ref{eq:qe}). The plot also
    presents the quadratic sum of the four contributions (solid black
    line) and the RMS of the 3 independent measurements recalled from
    the middle panel (dotted black line). }
  \label{fig:qe}
\end{figure*}

Our quantum efficiency measurement is obtained from Eq.~(\ref{eq:qe})
using the baseline gain estimate of \SI{1.269}{e^-/ADU} obtained in
Sect.~\ref{sec:DNL}. The measurement was performed three times and
then the optical configuration was varied to look for systematic
errors. The 3 independent measurements are presented in
Fig.~\ref{fig:qe}.

The largest systematic uncertainties affecting the overall scale of
the curve are i) the uncertainties in the gain determination and ii)
the $1$\% calibration uncertainty of the picoammeter reading the
photocurrent of the NIST reference sensor.\footnote{Quoting here the
  manufacturer calibration report.} Those uncertainties affect the
overall scale of the curve which therefore peaks at \SI{0.84\pm
  0.01}{e^-/ADU} for the measured position. The quantum efficiency
stays above $80\%$ in the range \SIrange{440}{570}{\nano\metre}, above
$50\%$ in the range \SIrange{387}{650}{\nm} and above $20\%$ all the
way up to \SI{790}{nm}.

The breakout of the noise contribution from the various components in
Eq.~(\ref{eq:qe}) to the statistical uncertainty is presented
in the bottom panel of Figure~\ref{fig:qe}, along with the
RMS of 3 measurements. The quadratic sum of the contributions nicely
lines up with the observed RMS of the measurements. The uncertainty is
dominated by background noise in the IMX411 sensor because a very
large aperture, matching exactly the spatial extent of the NIST
sensor, is used in order to minimize aperture correction
systematics. The statistical uncertainty on a single measurement with
nm resolution is smaller than \SI{0.001}{e^-/\gamma} over the majority of
the range and smaller \SI{0.01}{e^-/\gamma} everywhere.

\begin{figure*}
  \centering
  \includegraphics{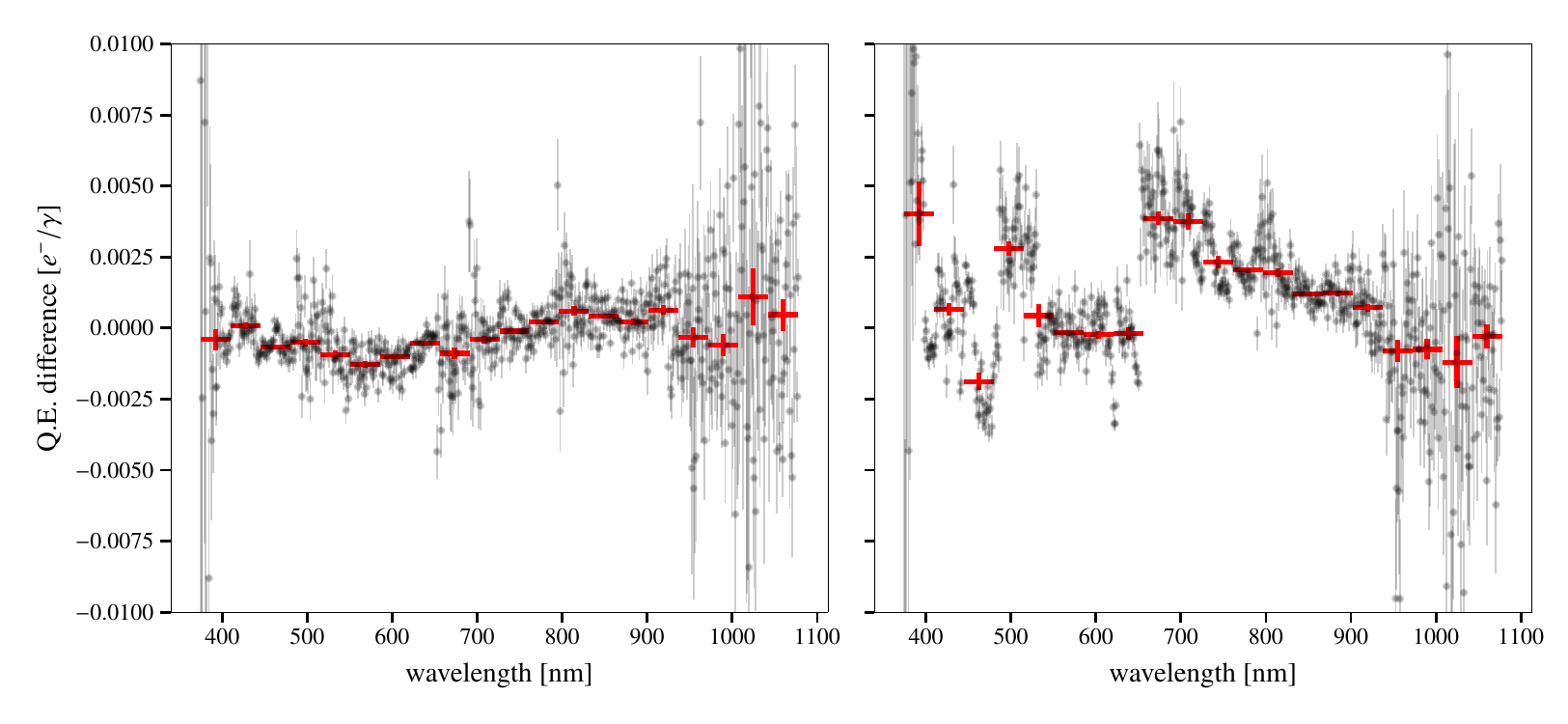}
  \caption{\emph{Left:} Difference between the baseline and stopped
    down measurement of the IMX411 quantum efficiency
    curve. \emph{Right:} Difference between two quantum efficiency
    measurements using 4s and 2s exposures.}
  \label{fig:syste}
\end{figure*}

Systematic uncertainties associated with the NIST calibration of the
reference photodiode are subdominant at this stage. We looked at two
other potential sources for systematic chromatic uncertainties. First,
a wavelength-dependent error would arise if the sensitive area of the
reference sensor and of the photometric aperture in the camera were
not perfectly matched, because the position and the shape of the
projected slit image changes slightly with wavelength. We tested this
possibility by stopping down the physical aperture in front of the
sensor by a factor of 2 in radius. This reduces the spatial extent of
the illuminated area on the sensor in all directions and makes the
measurement less sensitive to alignment issues. The difference between
the baseline and the stopped down measurement is presented on the left
panel in Fig.~\ref{fig:syste}. It suggests that the aperture mismatch
systematics are controlled at the \SI{0.001}{e^-/\gamma} level.

Finally, we looked at the potential impact of the non linearities on
the QE measurement. Non linearities may affect the measurement because
the illumination level is not constant across all wavelengths. We
tested this by doubling the camera exposure time and redoing the same
measurement. The difference between the two measurements is presented
on the right panel in Fig.~\ref{fig:syste}. The observed difference
reaches \SI{0.005}{e^-/\gamma} peak-to-peak, dominating the chromatic
uncertainty budget. In principle the effect of non-linearities could
be corrected but this correction was not attempted at this stage. A
specific measurement of the sensor integral linearity and exposure
time involving a precisely pulsed light source is planned for the
upgrade.

\section{Conclusion}
\label{sec:discussion}

The StarDICE sensor calibration bench presented above delivers quantum
efficiency measurements with statistical uncertainties below
\SI{e-3}{e^-/\gamma} in the range \SIrange{387}{950}{\nm}
(\SI{<e-2}{e^-/\gamma} in the range \SIrange{375}{1078}{\nm}) and low
systematic uncertainties (\SI{\sim e-3}{e^-/\gamma}) in quantum
efficiency ratios between different wavelengths, which is the metric
relevant for the measurement of type-Ia supernovae colors. The
measurement is sensitive to sensor non-linearities due to the change
in illumination intensity as a function of wavelength in the
monochromatic beam. A dedicated measurement of the sensor linearity is
therefore required to reach the requirements of StarDICE and the
addition of a tunable pulsed light source to the bench is planned for
this purpose. The systematic uncertainty on the gray-scale is
currently dominated by the calibration uncertainty of the picoammeter
providing the photocurrent of the reference photodiode.

The sensor calibration bench has been used for the calibration of the
quantum efficiency of a QHY411M camera hosting the \SI{150}{Mpixels}
IMX411ALR sensor from Sony. The camera was found to provide excellent
response stability (\SI{<0.0033}{\%} RMS in stable
conditions). Quantum efficiency is above 80\% in the range
\SIrange{440}{570}{\nm}, above \SI{50}{\%} on most of the visible
range, above \SI{20}{\%} all the way up to \SI{790}{\nm} and close to
\SI{10}{\%} at \SI{900}{nm}. This translates to an average quantum
efficiency in each of the photometric bands of the $ugrizy$ system of
respectively \SI{47}{\percent}, \SI{79}{\percent}, \SI{59}{\percent},
\SI{27}{\percent}, \SI{10}{\percent} and \SI{2}{\percent}. The
uniformity of the sensor response is at or below the \SI{1}{\%} RMS
level for all tested wavelengths. The sensor cosmetics is also
excellent. Degradation of the images from electrostatic influence
between pixels is expected to be negligible for most use cases given
the small measured size of the linear IPC and brighter-fatter
effect. This is especially appreciable for \SI{3.76}{\micro\metre}
pixel side. The only adverse effect found at this stage is a rather
large periodical differential non-linearity, despite apparently good
integral linearity. Such effects, commonly affecting ADCs,
  can require correction in applications sensitive to differential
  linearity, typically those measuring fluctuations around a variable
  level (see e.g. \citealt{2020A&A...641A...2P} for a successful
  account of \emph{a posteriori} corrections for ADC non-linearities
  in the context of the measurement of CMB fluctuations). A
two-parameter periodic alteration of the scale is sufficient to
mitigate the effect on the PTC up to \SI{\sim20}{ke^-}. More complex
alteration would be required to describe the entire scale, which was
not attempted. Also our integral linearity test relies on the accuracy
of the rolling shutter timing, which we did not control. We suggest
that applications sensitive to linearity may want to envision adequate
mapping of the scale.

We conclude that the sensor is very well suited for many astronomical
applications in the visible range. For the GRANDMA experiment, this
applies to the search of the optical counterparts of transitory events
with poor localization from gravitational wave or neutrino detections
which can be conducted in band $g$ and $r$. The characterization of
transients through their strong color evolution will however require
separated follow-up. Given the relatively low QE of the camera
(\SI{27}{\%}) in the $i$ band) longer exposures would be required to
reach adequate sensitivity for hunting kilonovae, necessitating
trade-off with survey size. Additionally, we note that the relatively
high QE in the $u$-band (\SI{47}{\%}) could be useful to capture the
earliest light of kilonovae in a particularly important band for
enabling model selection, provided that the source is detected early
enough.

The goal of the StarDICE experiment to control calibration systematic
uncertainties from \SI{375}{\nm} to \SI{1050}{\nm} at the \SI{0.1}{\%}
level appears difficult to reach in this first measurement due to the
low quantum efficiency beyond \SI{900}{\nm}. The StarDICE calibration
sensor bench is therefore being upgraded to increase the flux of the
input light-source in the infrared and to calibrate a
\SI{40}{\micro\metre} deep-depleted CCD sensor more sensitive at these
wavelengths. Other ongoing improvements include custom dust-protective
housing for the reference sensor, heat and temperature management,
upgrade of opto-mechanical parts to ease the alignment of the optics,
improved calibration of the reference sensor photocurrent reading and
dedicated measurement of the sensor linearity.

\subsection*{Acknowledgements}
\label{acknowledgements}
The authors thank the QHY company for providing the camera tested in
this study and for kindly answering our questions. Warm thanks are
also addressed to Pierre Astier and Pierre Antilogus for thoughtful
discussions and comments on DNL issues, and to Parker Fragelius and
Elana Urbach for their help in improving the text. This work received
support from the Programme National Cosmology et Galaxies (PNCG) of
CNRS/INSU with INP and IN2P3, co-funded by CEA and CNES and from the
DIM ACAV program of the Île-de-France region.

\bibliographystyle{aa}
\bibliography{biblio}

\appendix
\section{Statistical model of the photon transfer curve}
\label{sec:stat-model-phot}

\subsection{Simplest model, perfect illumination, perfect linearity}
\label{sec:simple}

Let's first consider an oversimplified model, where we assume
independence between successive images, a simple model of white
Gaussian readout noise, perfect linearity of the readout chain and
ignore potential electrodynamics effects such as the so-called
brighter-fatter effect strongly affecting the shape of the PTC in
thick deep-depleted CCDs.

The ADU counts in pixel $p \in [1, \cdots, P]$ and image
$i \in [1, \cdots, I]$ is then:
\begin{equation}
  \label{eq:model}
  \pixcount = \frac{N_{p,i} + n_{p,i}}{G_p} + b_p
\end{equation}
where $N_{p, i}\sim \mathcal{P}(\bar N_{p,i})$ is the number of
photoelectron converted and stored in pixel $p$ expected to follow a
Poisson distribution, $G_p$ is the effective gain of the readout for
the pixel (in \si{e^-/ADU}),
$n_{p,i} \sim \mathcal{N}(0, \sigma_p^2)$ is the noise in the
readout expressed in $e^-$, and $b_p$ is the tunable bias of the
readout electronic in \si{ADU}. The statistics defined in equations
(\ref{eq:stat}-\ref{eq:statlast}) have the following properties:
\begin{eqnarray}
  \label{eq:exp}
  \E [M_p] &=& \frac1{IG_p}\sum_i \bar N_{p,i} + b_p \\
  \var [M_p] &=& \frac1{I^2G^2_p}\sum_i \bar N_{p,i} + \frac{\sigma_p^2}{G^2I} \\
  \E [M_p^2] &=& \E[M_p]^2 + \var[M_p]\\
  \E [V_p^0] &=& \frac1{I}\sum_i \var[\pixcount] + \E [\pixcount]^2\\
           &=& \frac{\sigma_p^2}{G_p^2} + \frac{\sum_i \bar N_{pi}}{IG_p^2}  + \frac{\sum_i \left(\frac{\bar N_{p,i}}{G_p}+b_p\right)^2}{I}
\end{eqnarray}
Under perfect stability of the illumination conditions, that is $\bar{N}_{p,i} = \bar{N}_p$, the above simplify as:
\begin{eqnarray}
  \label{eq:simple}
  \E [M_p] &=& \frac{\bar N_p}{G_p} + b_p\\
  \var [M_p] &=& \frac{\bar N_p}{IG_p^2} + \frac{\sigma_p^2}{IG_p^2}\\
  \E [V_p^0 - M_p^2] &=& \frac{\sigma_p^2}{G_p^2} + \frac{\bar N_p}{G_p^2}\\
\end{eqnarray}
\begin{equation}
  \label{eq:result}
  \E[V_p^0 - M_p^2] = \frac{I-1}{I}\left(\frac{\E[M_p]-b}{G_p} + \frac{\sigma_p^2}{G_p^2}\right)
\end{equation}
In this simple model, the slope of the relation between the two
observables $V^0_p - M_p^2$ and $M_p$ gives a direct estimate of the
effective readout gain, while the intercept is a measurement of the
read noise.

The uncertainty on the determination of $G$ scales as $2/\sqrt{I}$,
which indicates that a determination of $G_p$ at $1$\% would require
of the order of $3\cdot 10^4$ images.

\subsection{Accommodating illumination variations}
\label{sec:variation}

The model can easily be extended to accommodate small fluctuations of
the illumination. Let us write $\bar N_{p,i} = \bar N_p(1 + \Delta_i)$
where we take by definition $\sum_{i \in I} \Delta_i = 0$. Denoting
$\Delta^2 = \frac{\sum\Delta_i^2}{I}$, the expectation of $V^0_p$ is
modified as:
\begin{equation}
  \E [V^0_p] = \frac{\bar N_p}{G_p^2}
  + \left(\frac{\bar N_p}{G_p} + b_p\right)^2
  +  \frac{\sigma_p^2}{G_p^2}
  +\Delta^2 \frac{\bar N_p^2}{G_p^2}
\end{equation}
and therefore the relation becomes:
\begin{equation}
  \label{eq:resultillu}
  \E[V_p - M_p^2] = \frac{I-1}{I}\left(\frac{\E[M_p]-b_p}{G_p} + \frac{\sigma_p^2}{G_p^2}\right) + \Delta^2(\E[M_p] - b_p)^2
\end{equation}
Close to the full-well $(\E[M_p - b_p] \sim 5 \cdot 10^5)$,
illumination fluctuations of the order of $10^{-4}$ would contribute a
bias of the order of \SI{0.5}{\%}, which may deserve some compensation
to investigate non-linearity at this level of accuracy. As
fluctuations of this order of magnitude are occasionally observed, we
choose to estimate $\Delta$ from the empirical variance of the third
observable $m_i$. Specifically denoting $m = \frac1N \sum_i m_i$,
$v = \frac1N\sum_i m_i^2$,
$\bar N = \frac1P \sum\frac{\bar N_p}{G_p}$, $B = \frac1P \sum b_p$,
$\sigma^2 = \frac1P \sum\frac{\sigma_p^2}{G_p}$, and
$G= \frac{\sum \bar N_p / G_p}{\sum \bar N_p/G_p^2}$ we have that:
\begin{equation}
  \label{eq:empirical}
  \E[v - m^2] = \frac{I-1}{I}\left(\frac{\bar N}{P G} + \frac{\sigma^2}{P}\right) + \Delta^2 \bar N^2
\end{equation}
and we form:
\begin{equation}
  \label{eq:delta}
  \hat \Delta^2 = \frac{v - m^2 - (I-1)\sigma^2/(PI)}{(m-B)^2}
\end{equation}

The estimate $\hat \Delta$ is used to subtract the contribution to
Eq.~(\ref{eq:empirical}) from reported statistics.

\subsection{Probability mass function of pixels with variable bit size}
\label{sec:variablebit}

The most striking feature of the readout chain we tested is a
periodical differential non-linearity, well reproducible across the
focal plane. Although the details of the readout chain are not known to
us, this could be interpreted as inaccurate bit widths affecting,
\emph{e.g.} a shared ramp in a ramp-compare ADC scheme. We suggest to
handle such effects by numerical computation of the probability mass
function for the pixel value. Let us denote $B$ the resolution of the
ADC. We then define for $n\in{1,\cdots, 2^{B}-1}$,
$D_n = \sum_{k=0}^{Nbits-1} (1+\delta_k) (2^k \land n)$ the successive
values generated by the ramp DAC, where $\land$ denote the binary
\emph{and} operator, and $\delta_k$ is the small error on the size of
the $k^{th}$ DAC bit. Furthermore we pose $D_0 = -\infty$ and
$D_{2^B} = \infty$. The probability mass function for the value
of the pixel $P(p=n)$ (dropping indices for clarity) is 0 for $n<0$ and $n \geq 2^B$ and:
\begin{align}
  \label{eq:pmf}
  P(p=n) &= P\left(D_n \leq \frac{N}{G} + n < D_{n+1}\right)\\
         &= \sum_{k=0}^\infty P(N=k)P\left(D_n - \frac{k}{G} < n < D_{n+1} - \frac{k}{G}\right)\\
         &= \sum_{k=0}^{\infty} \frac{\bar N^k e^{-\bar N}}{k!} \frac12 \left[\erf\left(\frac{D_{n+1} - \frac{k}{G} - b}{\sqrt2 \sigma}\right)
             -\erf\left(\frac{D_{n} - \frac{k}{G} - b}{\sqrt2 \sigma}\right)\right]
\end{align}
from which we obtain the expectation and variance of the pixel value
for a given illumination. In practice, the above can be truncated into
the product of finite matrices
$P = R(G, b, \sigma, \delta) \otimes Q(\bar N)$. We apply the
truncation $Q_k(\bar N) = \frac{\bar N^k e^{-\bar N}}{k!}$ for
$ \mathrm{min}(0, \bar N - 5\sqrt{\bar N})\leq k <\bar N + 5\sqrt{\bar
  N} + 4$ and $0$ elsewhere. Similarly zeroing $R_{n,k}$ for
$\vert n - \frac{k}{G} - b \vert > 6\sigma$ adequately speed up the
computation. We provide a python code using sparse matrix algebra to
compute the probability mass function of the pixels with arbitrary
digital boundaries and the observable statistics.

\subsection{Brighter-fatter effect}
\label{sec:bf}

Effects like the brighter-fatter could alter the above statistics. As
the probability of having a photon converted in a given pixel depends
on the charge already accumulated in the pixel, $\pixcount$ is no
longer Poisson distributed. A first order model of the pdf can be
built assuming that the shrinkage of the pixel area (or for what
matters of the depletion region in CIS) is proportional to the
fluctuation of the charge above the mean, with proportionality
constant $a$.
\begin{equation}
  \label{eq:Ik}
  I_k(t) = I\left(1+a(\E[N(t)] - k)\right)
\end{equation}
For the integration the effect needs to saturate in some way, as the
equation above would lead to non physical values. The saturation of the
tested sensor ensure that $k\lessapprox \SI{70}{\kilo e^-}$, so that
the equation will remain well behaved on the useful range for the
anticipated value of the proportionality constant $a \sim
10^{-7}$. Let us denote $P_k(t) = P(N(t) = k)$ the probability of
counting $k$ electrons in the pixel at time $t$. Counting events obeys
the following conditional probabilities:

\begin{eqnarray}
  \label{eq:diffeq}
  P\left[N(t+h)=k+1 | N(t)=k\right] &=& h I_k(t) + o(h)\\
  P\left[N(t+h)=k+1 | N(t)=k+1\right] &=&1 -h I_{k+1}(t) + o(h)
\end{eqnarray}
from which we deduce the differential equation verified by $P_{k}(t)$:
\begin{eqnarray}
  \label{eq:diffbf}
  \dot P_0 &=& -I_0P_0\\
  \dot P_k &=& I_{k-1} P_{k-1} - I_{k}P_{k} \quad\forall k, 0<k<K\\
  \dot P_K &=& I_{K-1}P_{K-1}
\end{eqnarray}

A numerical integration of the system shows that all this is very well
approximated by a Gaussian distribution with appropriate variance
given by \cite[Eq. 16]{astier}. As a shortcut, the Gaussian
approximation with modified variance is used in place of the Poisson
distribution to compute matrix $Q$ in equation~(\ref{eq:pmf}) when
brighter-fatter effect is taken into account in the PTC model.

\end{document}